\documentclass[prd,aps,showpacs,epsf,floats,10pt,onecolumn]{revtex4}%
\usepackage{amssymb}
\usepackage{amsfonts}
\usepackage{amsmath}
\usepackage{graphicx}%
\providecommand{\U}[1]{\protect\rule{.1in}{.1in}}

\begin{document}
\title{\textbf{Theoretical investigations of an information geometric approach to
complexity}\ }
\author{\textbf{Sean Alan Ali}$^{1}$ and \textbf{Carlo Cafaro}$^{2}$}
\affiliation{$^{1}$Albany College of Pharmacy and Health Sciences, 12208 Albany, New York, USA}
\affiliation{$^{2}$SUNY Polytechnic Institute, 12203 Albany, New York, USA}

\begin{abstract}
It is known that statistical model selection as well as identification of
dynamical equations from available data are both very challenging tasks.
Physical systems behave according to their underlying dynamical equations
which, in turn, can be identified from experimental data. Explaining data
requires selecting mathematical models that best capture the data
regularities. The existence of fundamental links among physical systems,
dynamical equations, experimental data and statistical modeling motivate us to
present in this article our theoretical modeling scheme which combines
information geometry and inductive inference methods to provide a
probabilistic description of complex systems in the presence of limited
information. Special focus is devoted to describe the role of our entropic
information geometric complexity measure. In particular, we provide several
illustrative examples wherein our modeling scheme is used to infer macroscopic
predictions when only partial knowledge of the microscopic nature of a given
system is available. Finally, limitations, possible improvements, and future
investigations are discussed.

\end{abstract}

\maketitle

\bigskip\pagebreak

\section{Introduction}

The inherent relationships among dynamics, modeling and complexity is indeed a
remarkable occurrence in the physical sciences \cite{jimmy}. In actual
experiments, information associated with the state of a physical system is
measured and collected at various points in space and time. In order to obtain
an understanding of the physics underlying the behavior of the system, the
dynamical equations governing the evolution of the system must be
reconstructed from the data. Indeed, the deduction of dynamical laws from
empirical data is a fundamental aspect of science \cite{minghia1, vijay,
minghia2}. In the recent work \cite{toby}, it was shown that the deduction of
the dynamical equations of a system from empirical data is NP hard and
computationally intractable. Moreover, this result is valid for both classical
and quantum systems, independent of the amount of data that is collected. This
seems to suggest that various intimately related issues, such as the
determination of dynamical equations that best approximates data, or comparing
families of dynamical models to data are generally intractable.

It is known that analysis of a system's data can, in some cases, enable
identification of classes of regularities in the behavior of relevant
variable(s) of the actual system. It is widely accepted that classes
exhibiting either complete regularity (i.e. perfectly ordered) or nonexistent
regularity (i.e. maximally random) would be classified as entirely non-complex
in the sense that such systems possess no structure \cite{gell-mann, james1,
james2}. By contrast, structured systems that admit correlations among the
constituents of the system can be very complex. Indeed, correlation and
structure are not entirely independent of randomness.

Well defined, useful measures of complexity are generally introduced in
scenarios that take into consideration the complete sequence of events that
lead to the emergence of the system whose complexity is being quantified
\cite{landauer}. For such complexity measures, only those states which are
reached through a difficult sequence of intermediate states is deemed complex.
For example, the notion of pattern is important in quantifying the complexity
of a noisy quantum channel \cite{romano}. The logical and thermodynamic depths
also play the role of complexity measures. The thermodynamic depth was
proposed by Lloyd and Pagels and represents the amount of entropy produced
during the evolution of a state of a system \cite{lloyd1}. The logical depth
was proposed by Bennett and represents the run time needed for a universal
Turing machine to execute the minimal program that reproduces (for example) a
system's configuration \cite{bennett-f}.

Since the sequence of intermediate states leading to the final state of a
system is of primary importance when defining a proper measure of complexity,
application of simple thermodynamic criteria to the states being compared are
generally inadequate. For instance, thermodynamic potentials adequately serve
to measure a system's capacity for irreversible change, but do not agree with
accepted notions of complexity \cite{bennett-book}. Specifically, the
thermodynamic entropy is a monotone functional of temperature in which high
(low) temperature corresponds to high (low) randomness, respectively. However,
since there exist multiple functions which vanish in the regimes of extreme
order and disorder, it is evident that this property alone does not adequately
constrain a useful statistical complexity measure \cite{james2}. By
statistical complexity measure we mean a quantity that measures the average
quantity of memory required to statistically reproduce a given configuration.
Despite these facts however, it is undeniable that thermodynamics plays a
critical role in the characterization and understanding of the complexity in
reversible, dissipative systems \cite{bennett-f}.

The difficulty encountered in the construction of a viable theory from a given
set of data can be approximately identified with the notion of cripticity. On
the other hand, the difficulty associated with extracting predictions from a
theory can be viewed as a loose interpretation of the concept of logical
depth. Both cripticity and logical depth are deeply related to complexity. The
ability to extract predictions from a theory can be quite difficult, but is
more so in composite systems involving interactions among subsystems. The
introduction of such interactions lead to the so-called fluctuation growth
which may give rise to nonlinear and chaotic dynamics. Such phenomena are
common and may occur in natural as well as artificial, complex dynamical
systems \cite{lloyd2013}. An issue of fundamental concern in the physics of
complex systems is model reduction. Model reduction in this context refers to
the identification of low-dimensional models that capture the gross features
of the original high-dimensional system \cite{jie}. It is often true that
consideration of the dynamics of a system alone may not be sufficient to
render reliable predictions. In such cases, entropic considerations must be
accounted for as well \cite{jaynes1}.

In this article, we make use of the so-called Entropic Dynamics (ED)
\cite{caticha-ED}, a theoretical framework built on both Maximum relative
Entropy (MrE) methods \cite{caticha-giffin} and information geometric
techniques \cite{amari-japan}. We emphasize that ED is formally similar to
other generally covariant theories: the dynamics is reversible, the
trajectories are geodesics, the system supplies its own notion of intrinsic
time, the motion can be derived from a variational principle of the form of
Jacobi's action principle rather than the more familiar principle of Hamilton
\cite{PR, chicone}. In brief, the canonical Hamiltonian formulation of ED is
an example of an information constrained dynamics where the
information-constraints play the role of generators of evolution. The ED
approach has been applied to the derivation of Newton's dynamics
\cite{caticha-cafaro} and various aspects of quantum theory \cite{caticha-jpa}%
. For more details on the ED, we refer to \cite{caticha-ED}.

Inspired by the ED approach to physics and motivated by the fundamental links
among physical systems, dynamical equations, experimental data and statistical
modeling, we present our theoretical information geometric scheme used to
model the dynamics of systems (of arbitrary nature) that are described by
probability distributions. We focus attention on the role of the our
information geometric complexity and entropy measures in characterizing
dynamical systems described by probability distributions. In particular, the
relationship among the information geometric entropy, the coarse-grained
Boltzmann, von Neumann and the Kolmogorov-Sinai (KS) dynamical entropies in
the appropriate regimes will be explored in our examples. The layout of this
article is as follows: in Section II, we review the MrE formalism used to
update probabilities with both information and data constraints. In Section
III, we present a brief introduction to Information Geometry (IG). In Section
IV, we introduce the information-geometric indicators of complexity for our
theoretical model, namely the information geometric entropy ($\mathcal{S}%
_{\mathcal{M}}$) and the information geometric complexity ($\mathcal{C}%
_{\mathcal{M}}$). In Section V, we present ten applications where our
theoretical model is used to study the dynamical complexity of statistical
models corresponding to the systems being investigated. Concluding remarks are
presented in Section VI.

\section{The Maximum relative Entropy Method}

The MrE method represents an efficient scheme for the updating of a family of
prior probability distribution functions when new information becomes
available in the form of constraints on the family of allowed posteriors. The
utility and versatility of the MrE method rests in the possibility of updating
a family of prior probability distribution functions in presence of both data
and constraints of the expected value type. This characteristic was initially
presented in a formal manner in \cite{caticha-giffin} wherein it was
demonstrated that Bayes updating can be viewed as a special case of the MrE
method.\ An analysis of the practical utility of this powerful feature of the
MrE method in real world applications was presented in \cite{giffin-caticha}.
We direct the reader to \cite{giffin16} for a recent application of the MrE
method to the physics of ferromagnetic materials.\textbf{ }In what follows, we
present the scheme for updating probabilities in the presence of both expected
value and data constraints.

In the remainder of this article, the quantities $x$ and $\theta$ represent
the microstate and macrostate of the system, respectively. Furthermore, the
microstates $x$ are elements of the microspace $\mathcal{X}$, while the
macrostates $\theta$ are elements of the parameter space $\mathcal{D}_{\theta
}$. We utilize the MrE method to update a prior probability distribution into
a posterior probability distribution. In\textbf{ }particular, we seek to
render inferences on some quantity $\theta\in\mathcal{D}_{\theta}$ given: i)
the prior information about quantity $\theta$; ii) the known functional
relationship among variables $x\in\mathcal{X}$ and $\theta\in\mathcal{D}%
_{\theta}$; and iii) the observed values of the quantity $x\in\mathcal{X}$.
The search space for the posterior probability distribution occurs within the
product space $\mathcal{X}\times\mathcal{D}_{\theta}$, while the joint
distribution is specified by $P\left(  x\text{, }\theta\right)  $. The
transition from $P_{\text{old}}\left(  \theta\right)  $ to $P_{\text{new}%
}\left(  \theta\right)  $ is stipulated by,%
\begin{equation}
P_{\text{new}}\left(  \theta\right)  \overset{\text{def}}{=}\int
dxP_{\text{new}}\left(  x\text{, }\theta\right)  \text{.}%
\end{equation}
The joint probability $P_{\text{new}}\left(  x\text{, }\theta\right)  $ serves
to maximizes the relative entropy $\mathcal{S}\left[  P\left\vert
P_{\text{old}}\right.  \right]  $\textbf{,}%
\begin{equation}
\mathcal{S}\left[  P\left\vert P_{\text{old}}\right.  \right]  \overset
{\text{def}}{=}-\int dxd\theta P\left(  x\text{, }\theta\right)  \log\left[
\frac{P\left(  x\text{, }\theta\right)  }{P_{\text{old}}\left(  x\text{,
}\theta\right)  }\right]  \text{,} \label{ef}%
\end{equation}
subject to the known information constraints. Observe that $P_{\text{old}%
}\left(  x\text{, }\theta\right)  $,%
\begin{equation}
P_{\text{old}}\left(  x\text{, }\theta\right)  =P_{\text{old}}\left(
x\left\vert \theta\right.  \right)  P_{\text{old}}\left(  \theta\right)
\text{,}%
\end{equation}
is known as the joint prior, while $P_{\text{old}}\left(  \theta\right)  $ and
$P_{\text{old}}\left(  x\left\vert \theta\right.  \right)  $ represent the
Bayesian prior and likelihood, respectively. It is worth noting that both the
joint prior and the standard Bayesian prior both serve to encode prior
information concerning the quantity $\theta\in\mathcal{D}_{\theta}$.

It should be noted at this juncture that the likelihood will be taken as prior
information due to its representation as the a priori established relation
between $\theta\in\mathcal{D}_{\theta}$ and $x\in\mathcal{X}$. The relevant
information constraints are enumerated as follows: first, we have the
normalization constraint,%
\begin{equation}
\int dxd\theta P\left(  x\text{, }\theta\right)  =1\text{.} \label{c1}%
\end{equation}
Second, we have the information constraint associated with some function
$f\left(  \theta\right)  $\textbf{ }specified by,%
\begin{equation}
\int dxd\theta f\left(  \theta\right)  P\left(  x\text{, }\theta\right)
\overset{\text{def}}{=}\left\langle f\left(  \theta\right)  \right\rangle
=F\text{.} \label{c2}%
\end{equation}
Finally, we are left to consider the observed data represented by $x^{\prime}%
$. In the context of the MrE framework, knowledge of this information
naturally leads to infinitely many constraints,%
\begin{equation}
\int d\theta P\left(  x\text{, }\theta\right)  =P\left(  x\right)
=\delta\left(  x-x^{\prime}\right)  \text{,} \label{c3}%
\end{equation}
for any $x\in\mathcal{X}$ where $\delta$ denotes the Dirac delta
function.\textbf{ }By means of the Lagrange multiplier technique, we proceed
to maximize the logarithmic relative entropy in Eq. (\ref{ef}) relative to the
constraints in Eqs. (\ref{c1}), (\ref{c2}), and (\ref{c3}). We
stipulate\textbf{ }that the variation of the entropy $\mathcal{S}\left[
P\left\vert P_{\text{old}}\right.  \right]  $ with respect to $P$ equals zero
identically,%
\begin{equation}
\delta\left\{
\begin{array}
[c]{c}%
\mathcal{S}\left[  P\left\vert P_{\text{old}}\right.  \right]  +\alpha\left[
\int dxd\theta P\left(  x\text{, }\theta\right)  -1\right] \\
+\beta\left[  \int dxd\theta f\left(  \theta\right)  P\left(  x\text{, }%
\theta\right)  -F\right] \\
+\int dx\gamma\left(  x\right)  \left[  \int d\theta P\left(  x\text{, }%
\theta\right)  -\delta\left(  x-x^{\prime}\right)  \right]
\end{array}
\right\}  =0\text{.} \label{var}%
\end{equation}
It is determined that Eq. (\ref{var}) can be reduced to\textbf{,}%
\begin{equation}
\int dxd\theta\left[  -\log P\left(  x\text{, }\theta\right)  -1+\log
P_{\text{old}}\left(  x\text{, }\theta\right)  +\alpha+\beta f\left(
\theta\right)  +\gamma\left(  x\right)  \right]  \delta P\left(  x\text{,
}\theta\right)  =0\text{,} \label{get}%
\end{equation}
for any $\delta P\left(  x\text{, }\theta\right)  $. Hence, from Eq.
(\ref{get}) we obtain%
\begin{equation}
P_{\text{new}}\left(  x\text{, }\theta\right)  =P_{\text{old}}\left(  x\text{,
}\theta\right)  \exp\left[  -1+\alpha+\beta f\left(  \theta\right)
+\gamma\left(  x\right)  \right]  \text{,} \label{newp}%
\end{equation}
where the Lagrange multipliers $\alpha$, $\beta$, and $\gamma\left(  x\right)
$ can be determined via substitution of Eq. (\ref{newp}) into Eqs. (\ref{c1}),
(\ref{c2}) and (\ref{c3}). After some algebraic manipulation, we are able to
obtain%
\begin{equation}
P_{\text{new}}\left(  x\text{, }\theta\right)  =\frac{\exp\left[  \beta
f\left(  \theta\right)  \right]  P_{\text{old}}\left(  x\text{, }%
\theta\right)  \delta\left(  x-x^{\prime}\right)  }{\int d\theta\exp\left[
\beta f\left(  \theta\right)  \right]  P_{\text{old}}\left(  x\text{, }%
\theta\right)  }\text{.}%
\end{equation}
Finally,\textbf{ }upon marginalizing $P_{\text{new}}\left(  x\text{, }%
\theta\right)  $ relative to the variable $x$, we deduce the desired updated
prior probability distribution function%
\begin{equation}
P_{\text{new}}\left(  \theta\right)  \overset{\text{def}}{=}\int
dxP_{\text{new}}\left(  x\text{, }\theta\right)  =\frac{\exp\left[  \beta
f\left(  \theta\right)  \right]  P_{\text{old}}\left(  x^{\prime}\text{,
}\theta\right)  }{\int d\theta\exp\left[  \beta f\left(  \theta\right)
\right]  P_{\text{old}}\left(  x^{\prime}\text{, }\theta\right)  }\text{.}
\label{gg}%
\end{equation}
For\textbf{ }convenience we define%
\begin{equation}
\Delta\left(  x^{\prime}\text{, }\beta\right)  \overset{\text{def}}{=}\int
d\theta\exp\left[  \beta f\left(  \theta\right)  \right]  P_{\text{old}%
}\left(  x^{\prime}\text{, }\theta\right)  \text{,}%
\end{equation}
in terms of which Eq. (\ref{gg}) becomes%
\begin{equation}
P_{\text{new}}\left(  \theta\right)  =P_{\text{old}}\left(  \theta\right)
P_{\text{old}}\left(  x^{\prime}\left\vert \theta\right.  \right)  \frac
{\exp\left[  \beta f\left(  \theta\right)  \right]  }{\Delta\left(  x^{\prime
}\text{, }\beta\right)  }\text{.} \label{ggg}%
\end{equation}
Observe that in scenarios lacking constraints of the form of expected values,
the parameter $\beta=0$ leading Eq. (\ref{ggg}) to reduce to the known Bayes
updating relation%
\begin{equation}
P_{\text{new}}\left(  \theta\right)  =\frac{P_{\text{old}}\left(
\theta\right)  P_{\text{old}}\left(  x^{\prime}\left\vert \theta\right.
\right)  }{P_{\text{old}}\left(  x^{\prime}\right)  }\text{.} \label{bb}%
\end{equation}
In an effort to be complete, we raise attention to the fact that Eq.
(\ref{bb}) may be obtained via combination of Bayes theorem,%
\begin{equation}
P_{\text{old}}\left(  \theta\left\vert x\right.  \right)  =\frac
{P_{\text{old}}\left(  \theta\right)  P_{\text{old}}\left(  x\left\vert
\theta\right.  \right)  }{P_{\text{old}}\left(  x\right)  }\text{,}%
\end{equation}
with Bayes rule,%
\begin{equation}
P_{\text{new}}\left(  \theta\right)  =P_{\text{old}}\left(  \theta\left\vert
x^{\prime}\right.  \right)  \text{.}%
\end{equation}

\section{Information Geometry}

In \cite{carlo-tesi}, the so-called Information Geometric Approach to Chaos
(IGAC) was presented. The IGAC makes use of the ED formalism to characterize
and quantify the complexity of geodesic information flows on statistical
manifolds underlying the entropic dynamics of physical systems described in
terms of probability distributions. The IGAC\ can be viewed as the information
geometric counterpart of conventional geometrodynamical approaches
\cite{casetti, di bari}, with the usual classical configuration space\ being
replaced with a suitable statistical manifold. In particular, the IGAC
represents an information geometric extension of the Jacobi geometrodynamics
\cite{jacobi}.

\subsection{The Fisher-Rao Information Metric}

In this subsection we describe some properties of the statistical manifold
$\mathcal{M}$. An $n$-dimensional $%
\mathbb{C}
^{\infty}$ differentiable manifold $\mathcal{M}$ can be viewed as a set of
points $p$ that admits coordinate charts $\mathcal{U}_{\mathcal{M}}$ such that
the following two conditions are satisfied: 1) each element $c\in
\mathcal{U}_{\mathcal{M}}$ is a one-to-one mapping from $\mathcal{M}$ to some
open subset of $%
\mathbb{R}
^{n}$; 2) given any one-to-one mapping $\phi$ from $\mathcal{M}$ to $%
\mathbb{R}
^{n}$ for all $c\in\mathcal{U}_{\mathcal{M}}$, the mapping $\phi\in
\mathcal{U}_{\mathcal{M}}\Leftrightarrow\phi\circ c^{-1}$ is a $%
\mathbb{C}
^{\infty}$ diffeomorphism. In particular, the set of probability distributions
$p\left(  x|\theta\right)  $ forms an $n$-dimensional statistical manifold
$\mathcal{M}$%
\begin{equation}
\mathcal{M}\overset{\text{def}}{=}\left\{  p\left(  x|\theta\right)
:\theta=\left(  \theta^{1}\text{,..., }\theta^{n}\right)  \in\mathcal{D}%
_{\theta}\right\}  \text{.}%
\end{equation}
The points\textbf{ }$p$\textbf{ }belonging to the space $\mathcal{M}$\ are
parameterized by $n$ real-valued variables $\left(  \theta^{1}\text{,...,
}\theta^{n}\right)  $. Hence, the parameters $\theta$ serve as coordinates for
the point $p$. The set $\mathcal{D}_{\mathbf{\theta}}$ is the whole parameter
space available to the system and is a subset of $%
\mathbb{R}
^{n}$%
\begin{equation}
\mathcal{D}_{\theta}\overset{\text{def}}{=}%
{\displaystyle\bigotimes\limits_{k=1}^{n}}
\mathcal{I}_{\theta^{k}}=\left(  \mathcal{I}_{\theta^{1}}\otimes
\mathcal{I}_{\theta^{2}}\text{...}\otimes\mathcal{I}_{\theta^{n}}\right)
\subseteq%
\mathbb{R}
^{n}\text{,} \label{int-space}%
\end{equation}
where $\mathcal{I}_{\theta^{k}}$ is a subset of $%
\mathbb{R}
$ and represents the full range of allowed values of the macrostates $\theta$.

At this point it is important to emphasize that there exists an infinite
number of Riemannian metrics on the manifold $\mathcal{M}$. For this reason, a
necessary and fundamental assumption within the IG framework is the choice of
the Fisher-Rao information metric as that which underlies the Riemannian
geometry of $\mathcal{M}$ \cite{amari-japan, fisher, rao}, namely%
\begin{equation}
g_{\mu\nu}\left(  \theta\right)  \overset{\text{def}}{=}\int dxp\left(
x|\theta\right)  \partial_{\mu}\log p\left(  x|\theta\right)  \partial_{\nu
}\log p\left(  x|\theta\right)  =4\int dx\partial_{\mu}\sqrt{p\left(
x|\theta\right)  }\partial_{\nu}\sqrt{p\left(  x|\theta\right)  }=-\left.
\left(  \frac{\partial^{2}\mathcal{S}\left(  \theta^{\prime}\text{, }%
\theta\right)  }{\partial\theta^{\prime\mu}\partial\theta^{\prime\nu}}\right)
\right\vert _{\theta^{\prime}=\theta}\text{,} \label{FRMetric}%
\end{equation}
with $\mu$, $\nu=1$,..., $n$ for an $n$-dimensional manifold, $\partial_{\mu
}\overset{\text{def}}{=}\frac{\partial}{\partial\theta^{\mu}}$ and
$\mathcal{S}\left(  \theta^{\prime}\text{, }\theta\right)  $ representing the
logarithmic relative entropy \cite{ARIEL},%
\begin{equation}
\mathcal{S}\left(  \theta^{\prime}\text{, }\theta\right)  =-\int dxp\left(
x|\theta^{\prime}\right)  \log\left[  \frac{p\left(  x|\theta^{\prime}\right)
}{p\left(  x|\theta\right)  }\right]  \text{.}%
\end{equation}
The information metric $g_{\mu\nu}\left(  \theta\right)  $ is a
positive-definite, symmetric Riemannian metric that defines a measure of
distinguishability among macrostates on $\mathcal{M}$. It satisfies the two
following properties: 1) invariance under (invertible) transformations of
elements of the microspace $\mathcal{X}$; 2) covariance under
reparametrization of the elements of the macrospace $\mathcal{D}_{\theta}$.
The invariance of $g_{\mu\nu}\left(  \theta\right)  $ under reparametrization
of the microspace $\mathcal{X}$ implies that \cite{amari-japan}\textbf{,}%
\begin{equation}%
\mathbb{R}
^{l}\subseteq\mathcal{X}\ni x\mapsto y\overset{\text{def}}{=}f\left(
x\right)  \in\mathcal{Y}\subseteq%
\mathbb{R}
^{l}\Longrightarrow p\left(  x|\theta\right)  \mapsto p^{\prime}\left(
y|\theta\right)  =\left[  \frac{1}{\left\vert \frac{\partial f}{\partial
x}\right\vert }p\left(  x|\theta\right)  \right]  _{x=f^{-1}\left(  y\right)
}\text{.}%
\end{equation}
The covariance under reparametrization of elements of\textbf{ }the parameter
space\textbf{\ }$\mathcal{D}_{\theta}$\textbf{\ }(homeomorphic to\textbf{\ }%
$\mathcal{M}$\textbf{) }implies that \cite{amari-japan},%
\begin{equation}
\mathcal{D}_{\theta}\ni\theta\mapsto\theta^{\prime}\overset{\text{def}}%
{=}f\left(  \theta\right)  \in\mathcal{D}_{\theta^{\prime}}\Longrightarrow
g_{\mu\nu}\left(  \theta\right)  \mapsto g_{\mu\nu}^{\prime}\left(
\theta^{\prime}\right)  =\left[  \frac{\partial\theta^{\alpha}}{\partial
\theta^{\prime\mu}}\frac{\partial\theta^{\beta}}{\partial\theta^{\prime\nu}%
}g_{\alpha\beta}\left(  \theta\right)  \right]  _{\theta=f^{-1}\left(
\theta^{\prime}\right)  }\text{,}%
\end{equation}
where $g_{\mu\nu}\left(  \theta\right)  $ is given in Eq\textbf{.}
(\ref{FRMetric}) and
\begin{equation}
g_{\mu\nu}^{\prime}\left(  \theta^{\prime}\right)  =\int dxp^{\prime}\left(
x|\theta^{\prime}\right)  \partial_{\mu}^{\prime}\log p^{\prime}\left(
x|\theta^{\prime}\right)  \partial_{\nu}\log p^{\prime}\left(  x|\theta
^{\prime}\right)  \text{,}%
\end{equation}
with $\partial_{\mu}^{\prime}=\frac{\partial}{\partial\theta^{\prime\mu}}$ and
$p^{\prime}\left(  x|\theta^{\prime}\right)  =p\left(  x|\theta=f^{-1}\left(
\theta^{\prime}\right)  \right)  $.

It is known from IG \cite{amari-japan} that there is a one-to-one relation
between elements of $\mathcal{M}$ and $\mathcal{D}_{\theta}$. In particular,
the statistical manifold $\mathcal{M}$ is homeomorphic to the parameter space
$\mathcal{D}_{\theta}$. This implies the existence of a continuous, bijective
map $h_{\mathcal{M},\mathcal{D}_{\theta}}$,%
\begin{equation}
h_{\mathcal{M},\mathcal{D}_{\theta}\text{ }}:\mathcal{M}\ni p\left(
x|\theta\right)  \mapsto\theta\in\mathcal{D}_{\theta},
\end{equation}
where $h_{\mathcal{M},\mathcal{D}_{\theta}\text{ }}^{-1}\left(  \theta\right)
=p\left(  x|\theta\right)  $.

\subsection{Entropic Motion}

Given a statistical manifold $\mathcal{M}$ with metric $g_{\mu\nu}\left(
\theta\right)  $, the ED program is concerned with the following issue
\cite{caticha-ED}: given the known initial $\theta_{\text{initial}}$ and final
$\theta_{\text{final}}$ states of a system, what is the expected trajectory of
the system as it evolves? In answering this question, the ED framework
implicitly assumes that a trajectory exists and furthermore, the trajectory
follows from a principle of inference, namely the MrE method
\cite{caticha-giffin}. These assumptions imply that large changes arise as a
consequence of a continuous succession of small incremental changes. By
considering only small changes in going from $\theta_{\text{initial}}$ to
$\theta_{\text{final}}=\theta_{\text{initial}}+\Delta\theta$, the distance
$\Delta l$ between such states is given by,%
\begin{equation}
\Delta l^{2}\overset{\text{def}}{=}g_{\mu\nu}\left(  \theta\right)
\Delta\theta^{\mu}\Delta\theta^{\nu}\text{.}%
\end{equation}
In what follows, we primarily follow the work of Caticha \cite{caticha-ED}. In
going from $\theta_{\text{initial}}$ to $\theta_{\text{final}}$, the system
must necessarily sample a midway point, namely a state $\theta$ that is
equidistant between $\theta_{\text{initial}}$ and $\theta_{\text{final}}$. We
remark that there is no special significance to this chosen halfway state. We
could equally well have argued that in evolving from $\theta_{\text{initial}}$
to $\theta_{\text{final}}$, the system must initially sample a state that is
doubly distant from $\theta_{\text{final}}$ as it is from $\theta
_{\text{initial}}$. More generally, it is necessary only for the system to
sample some intermediate state $\theta_{\xi}$ whereby having already covered a
distance $dl$ from $\theta_{\text{initial}}$, a distance $\xi dl$ alone
remains to be traversed in order to arrive at $\theta_{\text{final}}$. Midway
states correspond to $\xi$ $=1$, one-third of the way states correspond to
$\xi$ $=2$, and so on. Each value of $\xi$ provides a different criterion by
which the trajectory may be selected. If there exist a multiplicity of ways by
which a trajectory is determined, consistency demands all the different
approaches should agree. As such, the selected trajectory must necessarily be
independent of the choice of $\xi$.

In line with the above reasoning, the aforementioned issue of concern to the
ED program can be effectively re-framed as follows: if a system in initial
state $p(x|\theta_{\text{initial}})$ is subject to new information
constraints, then the system will evolve to one of the neighboring states in
the family $p(x|\theta_{\xi})$. The relevant question becomes: how do we
select the proper $p(x|\theta_{\xi})$? Such a reformulation of the ED problem
is naturally of the type that can be analyzed by means of the MrE method. In
particular, recall that the MrE method is designed for the specific purpose of
processing information. It enables the transition from an initial family of
models described by a prior probability distribution, to a new family of
models described by a posterior distribution when the available information is
represented by specification of the set of distributions from which the
posterior must be selected. Typically, this set of distributions is
constrained by the known moments of some relevant set of variables. This is
not necessary however, nor do the information-constraints need not be linear
functionals. In the ED framework, constraints are defined geometrically.

When using the MrE method, it is important to specify which entropy is to be
maximized. The selection of a distribution $p(x|\theta)$ requires maximization
of the entropy functional of form%
\begin{equation}
\mathcal{S}\left[  p|q\right]  \overset{\text{def}}{=}-\int dxp\left(
x|\theta\right)  \log\left[  \frac{p\left(  x|\theta\right)  }{q\left(
x\right)  }\right]  \text{.} \label{juve}%
\end{equation}
Equation (\ref{juve}) serves to define the entropy of $p\left(  x|\theta
\right)  $ relative to the prior $q(x)$. It is obvious that the selected
posterior distribution coincides with the prior distribution in the absence of
new information-constraints. Since the distribution that maximizes
$\mathcal{S}\left[  p|q\right]  $ when new information-constraints are lacking
is $p\propto q$, we must take $q(x)$ to represent the prior. Assuming the
system is known to be in the initial state $p(x|\theta_{\text{initial}})$ and
we do not have any information that the system has changed, then we have no
reason to assume any change has occurred. The prior $q(x)$ should therefore be
chosen such that maximization of $\mathcal{S}\left[  p|q\right]  $ subject to
no new information-constraints leads naturally to the posterior $q(x)=$
$p(x|\theta_{\text{initial}})$. By contrast, if the system is known to be in
the initial state $p(x|\theta_{\text{initial}})$ and we then obtain new
information that the system has evolved to one of the neighboring states
within the family $p(x|\theta_{\xi})$, then the correct selection of the
posterior probability distribution is obtained by maximizing the entropy,%
\begin{equation}
\mathcal{S}\left[  \theta|\theta_{\text{initial}}\right]  \overset{\text{def}%
}{=}-\int dxp(x|\theta)\log\left[  \frac{p(x|\theta)}{p(x|\theta
_{\text{initial}})}\right]  \text{,}%
\end{equation}
subject to the constraint $\theta=\theta_{\xi}$.

In order to facilitate analysis, we assume the system evolves from
$\theta_{\text{initial}}$ to $\theta_{\text{final}}=\theta_{\text{initial}%
}+\Delta\theta$. Moreover, we denote by $\theta_{\xi}$ $=$ $\theta
_{\text{initial}}$ $+d\theta$ with $\xi\in%
\mathbb{R}
_{0}^{+}$ an arbitrary intermediate state\textbf{ }that is infinitesimally
close to $\theta_{\text{initial}}$. Hence, the distance $d\left(
\theta_{\text{initial}}\text{, }\theta_{\text{final}}\right)  \overset
{\text{def}}{=}dl_{\text{initial}\rightarrow\text{final}}^{2}$ between
$\theta_{\text{initial}}$ and $\theta_{\text{final}}$ is given by
$dl_{\text{initial}\rightarrow\text{final}}^{2}=g_{\mu\nu}\left(
\theta\right)  \Delta\theta^{\mu}\Delta\theta^{\nu}$, while the distance
between $\theta_{\text{initial}}$ to and $\theta_{\xi}$ is given by,%
\begin{equation}
dl_{\text{initial}\rightarrow\xi}^{2}=g_{\mu\nu}\left(  \theta\right)
d\theta^{\mu}d\theta^{\nu}\text{.} \label{1}%
\end{equation}
The distance between $\theta_{\xi}$ and $\theta_{\text{final}}$ can be written
as,%
\begin{equation}
dl_{\xi\rightarrow\text{final}}^{2}=g_{\mu\nu}\left(  \theta\right)  \left(
\Delta\theta^{\mu}-d\theta^{\mu}\right)  \left(  \Delta\theta^{\nu}%
-d\theta^{\nu}\right)  \text{.} \label{2}%
\end{equation}
The MrE\ maximization problem then essentially reduces to that of maximization
of the functional%
\begin{equation}
S\left[  \theta_{\text{initial}}+d\theta|\theta_{\text{initial}}\right]
=-\frac{1}{2}g_{\mu\nu}\left(  \theta\right)  d\theta^{\mu}d\theta^{\nu
}=-\frac{1}{2}dl_{\text{initial}\rightarrow\xi}^{2}%
\end{equation}
under variations of $d\theta$ subject to the geometric constraint,%
\begin{equation}
\xi dl_{\text{initial}\rightarrow\xi}=dl_{\xi\rightarrow\text{final}}\text{.}
\label{constraint}%
\end{equation}
It necessarily follows that%
\begin{equation}
\delta\left[  -\frac{1}{2}g_{\mu\nu}\left(  \theta\right)  d\theta^{\mu
}d\theta^{\nu}-\gamma\left(  \xi^{2}dl_{\text{initial}\rightarrow\xi}%
^{2}-dl_{\xi\rightarrow\text{final}}^{2}\right)  \right]  =0\text{,} \label{3}%
\end{equation}
where $\gamma$ denotes a Lagrangian multiplier. Substituting Eqs. (\ref{1})
and (\ref{2}) into Eq. (\ref{3}), we find%
\begin{equation}
\left\{  \left[  1+2\gamma\left(  \xi^{2}-1\right)  \right]  d\theta_{\mu
}+2\gamma\Delta\theta_{\mu}\right\}  \delta\left(  d\theta^{\mu}\right)
=0\text{.} \label{4}%
\end{equation}
Since Eq. (\ref{4}) must be satisfied for any $\delta\left(  d\theta^{\mu
}\right)  $, it must be be true that $\left\{  \left[  1+2\gamma\left(
\xi^{2}-1\right)  \right]  d\theta_{\mu}+2\gamma\Delta\theta_{\mu}\right\}
=0$, that is,%
\begin{equation}
d\theta_{\mu}=\chi\Delta\theta_{\mu}\text{,} \label{5}%
\end{equation}
where $\chi=\chi\left(  \xi\text{, }\gamma\right)  $ is defined as,%
\begin{equation}
\chi\left(  \xi\text{, }\gamma\right)  \overset{\text{def}}{=}\frac{1}{\left(
1-\xi^{2}\right)  -\frac{1}{2\gamma}}\text{.} \label{6}%
\end{equation}
In order to determine the value of the Lagrange multiplier $\gamma$, we
recognize that the geometric constraint in Eq. (\ref{constraint}) can be
recast as, $\xi^{2}dl_{\text{initial}\rightarrow\xi}^{2}-dl_{\xi
\rightarrow\text{final}}^{2}=0$. Then, by using Eqs. (\ref{1}), (\ref{2}) and
(\ref{5}), we are able to\textbf{ }obtain $\left[  \xi^{2}\chi^{2}-\left(
1-\chi\right)  ^{2}\right]  g_{\mu\nu}\left(  \theta\right)  \Delta\theta
^{\mu}\Delta\theta^{\nu}=0$, such that%
\begin{equation}
\xi^{2}\chi^{2}-\left(  1-\chi\right)  ^{2}=0\text{.} \label{7}%
\end{equation}
Combination of Eqs. (\ref{6}) and (\ref{7}), leads to%
\begin{equation}
\chi\left(  \xi\right)  \overset{\text{def}}{=}\frac{1}{1+\xi}\text{ and,
}\gamma\left(  \xi\right)  \overset{\text{def}}{=}-\frac{1}{2\xi\left(
1+\xi\right)  }\text{.}%
\end{equation}
In this manner we were able to determine%
\begin{equation}
dl_{\text{initial}\rightarrow\xi}^{2}\overset{\text{def}}{=}\frac{1}{\left(
1+\xi\right)  ^{2}}\Delta\theta^{2}\text{,} \label{8}%
\end{equation}
and%
\begin{equation}
dl_{\xi\rightarrow\text{final}}^{2}\overset{\text{def}}{=}\frac{\xi^{2}%
}{\left(  1+\xi\right)  ^{2}}\Delta\theta^{2}\text{.} \label{9}%
\end{equation}
From Eqs. (\ref{8}) and (\ref{9}), it is true that%
\begin{equation}
dl_{\text{initial}\rightarrow\xi}+dl_{\xi\rightarrow\text{final}}=\frac
{1}{1+\xi}\Delta\theta\text{ }+\frac{\xi}{1+\xi}\Delta\theta=\Delta
\theta\text{.} \label{10}%
\end{equation}
However, since $dl_{\text{initial}\rightarrow\text{final}}^{2}\overset
{\text{def}}{=}g_{\mu\nu}\left(  \theta\right)  \Delta\theta^{\mu}\Delta
\theta^{\nu}=\Delta\theta^{2}$, we have%
\begin{equation}
dl_{\text{initial}\rightarrow\text{final}}=\Delta\theta\text{.} \label{11}%
\end{equation}
By combining Eqs. (\ref{10}) and (\ref{11}), we are able to show that
$dl_{\text{initial}\rightarrow\text{final}}=dl_{\text{initial}\rightarrow\xi
}+dl_{\xi\rightarrow\text{final}}$. Thus, given
\begin{equation}
\Delta\theta\overset{\text{def}}{=}d\theta+\left(  \Delta\theta-d\theta
\right)  \text{,} \label{12}%
\end{equation}
we have determined that,%
\begin{equation}
\left\Vert \Delta\theta\right\Vert =\left\Vert d\theta\right\Vert +\left\Vert
\Delta\theta-d\theta\right\Vert \text{,} \label{13}%
\end{equation}
where $\left\Vert \Delta\theta\right\Vert \overset{\text{def}}{=}%
\sqrt{dl_{\text{initial}\rightarrow\text{final}}^{2}}$, $\left\Vert
d\theta\right\Vert \overset{\text{def}}{=}\sqrt{dl_{\text{initial}%
\rightarrow\xi}^{2}}$ and, $\left\Vert \Delta\theta-d\theta\right\Vert
\overset{\text{def}}{=}\sqrt{dl_{\xi\rightarrow\text{final}}^{2}}$. In view of
Eq. (\ref{12}), we conclude that Eq. (\ref{13}) is satisfied provided that
$d\theta$ and $\Delta\theta-d\theta$ are collinear. That is to say, the
triangle defined by the triple $\theta_{\text{initial}}$, $\theta_{\xi}$,
$\theta_{\text{final}}$ will degenerate into a straight line. This result is
sufficient to determine a short segment of the trajectory since all
intermediate states lie on the straight line whose endpoints are
$\theta_{\text{initial}}$ and $\theta_{\text{final}}$. The generalization
beyond short trajectories is as follows: if any three neighboring points along
a curve lie along a straight line segment, then the curve in question is
necessarily a \emph{geodesic}. We emphasize that this result is independent of
the arbitrarily chosen $\xi$.

A geodesic on an $n$-dimensional statistical manifold $\mathcal{M}$ represents
the maximum probability path a complex dynamical system explores in its
evolution from $\theta_{\text{initial}}$ to $\theta_{\text{final}}$. Each
point that lies on the geodesic curve is parametrized by the
variables\textbf{\ }$\theta$ defining the macrostates of the system. Each
component $\theta^{\kappa}$ with $\kappa=1$,..., $n$ is a solution of the
geodesic equation \cite{caticha-ED, defelice, carmo},
\begin{equation}
\frac{d^{2}\theta^{\kappa}}{d\tau^{2}}+\Gamma_{\mu\nu}^{\kappa}\frac
{d\theta^{\mu}}{d\tau}\frac{d\theta^{\nu}}{d\tau}=0\text{.} \label{geo1}%
\end{equation}
Furthermore, each macrostate $\theta$ is in a one-to-one correspondence with
the probability distribution $p\left(  x|\theta\right)  $ representing the
maximally probable description of the system being considered. This is a
distribution of the microstates $x$. In summary, the solution to the main ED
problem is as follows \cite{caticha-ED}: \emph{the expected trajectory between
known initial and final states is the geodesic curve that connects them}. It
is worth noting however, that this result alone is insufficient to conclude
that the actual trajectory coincides with the expected trajectory. An
affirmative result to this issue depends on whether the information encoded in
the initial state is sufficient for such prediction. For further discussion on
this particular issue, we refer to Ref. \cite{carlopre16}.

\subsection{Volumes in Curved Statistical Manifolds}

Upon establishing a quantitative notion of distinguishability among
probability distributions in terms of distances assigned relative to the
Fisher-Rao information metric $g_{\mu\nu}\left(  \theta\right)  $, we may
construct the Riemannian volume element $d\mathcal{V}_{\mathcal{M}}$ to be
used as the natural measure in the space of distributions. To this end, we
consider an $n$-dimensional volume of the statistical manifold $\mathcal{M}$
of distributions $p\left(  x|\theta\right)  $ labelled by parameters
$\theta^{\kappa}$ with $\kappa=1$,..., $n$. The parameters $\theta^{\kappa}$
represent coordinates for the point $p$.\textbf{ }Furthermore, we consider
very small regions of the manifold $\mathcal{M}$ wherein we employ\textbf{
}Cartesian coordinates where the metric assumes\ the form of the identity
matrix $\delta_{ab}$ since curved spaces are locally flat. In locally
Cartesian coordinates $\varphi$, the volume element is given by the product
$d\mathcal{V}_{\mathcal{M}}\overset{\text{def}}{=}%
{\displaystyle\prod\limits_{k=1}^{n}}
d\varphi^{k}$, which in terms of the old coordinates $\theta^{\kappa}$ reads,%
\begin{equation}
d\mathcal{V}_{\mathcal{M}}\overset{\text{def}}{=}\left\vert \frac
{\partial\varphi}{\partial\theta}\right\vert d\theta^{1}d\theta^{2}\text{...
}d\theta^{n}\text{.}%
\end{equation}
The next clear task is the\textbf{ }evaluation of the Jacobian $\left\vert
\frac{\partial\varphi}{\partial\theta}\right\vert $ associated with the
transformation that takes the metric $g_{\mu\nu}$ into its flat (i.e.
Euclidean) form $\delta_{ab}$. We define the new coordinates as $\varphi
^{a}\overset{\text{def}}{=}\Phi^{a}\left(  \theta^{1}\text{,...., }\theta
^{n}\right)  $ where $\Phi$ denotes a coordinate transformation map
$\Phi:\theta\rightarrow\varphi$. Therefore, a small change $d\theta$
corresponds to a small change $d\varphi$ according to%
\begin{equation}
d\varphi^{a}\overset{\text{def}}{=}X_{\mu}^{a}d\theta^{\mu}\text{ where
}X_{\mu}^{a}\overset{\text{def}}{=}\frac{\partial\varphi^{a}}{\partial
\theta^{\mu}}\text{,}%
\end{equation}
and the Jacobian is given by the determinant of the matrix $X_{\mu}^{a}$,
$\left\vert \frac{\partial\varphi}{\partial\theta}\right\vert \overset
{\text{def}}{=}\left\vert \det\left(  X_{\mu}^{a}\right)  \right\vert $. The
distance between two neighboring points is the same whether it is computed in
either the old or new coordinates, $dl^{2}=g_{\mu\nu}d\theta^{\mu}d\theta
^{\nu}=\delta_{ab}d\varphi^{a}d\varphi^{b}$. Hence, the old and the new
metrics are related via $g_{\mu\nu}\left(  \theta\right)  =\delta_{ab}X_{\mu
}^{a}X_{\nu}^{b}$.

We now consider statistical manifolds with Fisher-Rao information metric
tensor $g_{\mu\nu}\left(  \theta\right)  $ with $\theta\overset{\text{def}}%
{=}\left(  \theta^{1}\text{,..., }\theta^{n}\right)  $ whose determinant can
be written as,%
\begin{equation}
\det\left[  g_{\mu\nu}\left(  \theta\right)  \right]  =g\left(  \theta
^{1}\text{,..., }\theta^{n}\right)  =%
{\displaystyle\prod\limits_{\kappa=1}^{n}}
g_{\kappa}\left(  \theta^{\kappa}\right)  \text{.} \label{uno}%
\end{equation}
From Eq. (\ref{uno}), we assume that the determinant $g\left(  \theta\right)
$ can be factorized in a product of $n$-functions $g_{\kappa}\left(
\theta^{\kappa}\right)  $ with $1\leq\kappa\leq n$, where each function
depends, at worse, on a single variable $\theta^{\kappa}$. If there is no
dependence on the variable $\theta^{\kappa}$, we simply assign $g_{\kappa
}\left(  \theta^{\kappa}\right)  =1$. Note that both uncorrelated and
correlated Gaussian statistical manifolds satisfy this set of assumptions. If
Eq. (\ref{uno}) is not satisfied however, it is only required that
$\sqrt{g\left(  \theta\right)  }$ to be integrable over the $n$-dimensional
space $\mathcal{D}_{\mathbf{\theta}}$ defined in Eq. (\ref{int-space}). Thus,
upon taking the determinant of $g_{\mu\nu}\left(  \theta\right)  $, we obtain
$g\left(  \theta\right)  \overset{\text{def}}{=}\det\left[  g_{\mu\nu}\left(
\theta\right)  \right]  =\left[  \det\left(  X_{\mu}^{a}\right)  \right]
^{2}$ and therefore $\left\vert \det\left(  X_{\mu}^{a}\right)  \right\vert
=\eta_{\left(  \mathcal{M}\text{, }g\right)  }\left(  \theta^{1}\text{,...,
}\theta^{n}\right)  $, where $\eta_{\left(  \mathcal{M}\text{, }g\right)
}\left(  \theta^{1}\text{,..., }\theta^{n}\right)  $ is the so-called Fisher
density and equals the square root of the determinant of the metric tensor
$g_{\mu\nu}\left(  \theta\right)  $
\begin{equation}
\eta_{\left(  \mathcal{M}\text{, }g\right)  }\left(  \theta^{1}\text{,...,
}\theta^{n}\right)  \overset{\text{def}}{=}\sqrt{g\left(  \theta\right)
}\text{.}%
\end{equation}
Observe that $\sqrt{g\left(  \theta\right)  }d^{n}\theta$ is a scalar quantity
and is consequently invariant under orientation preserving \textbf{(}i.e. with
positive Jacobian\textbf{)} general coordinate transformations $\theta
\rightarrow\theta^{\prime}$. The square root of the determinant $g\left(
\theta\right)  $ of the metric tensor $g_{\mu\nu}\left(  \theta\right)  $ and
the flat infinitesimal volume element $d^{n}\theta$ transform as,%
\begin{equation}
\sqrt{g\left(  \theta\right)  }\overset{\theta\rightarrow\theta^{\prime}%
}{\rightarrow}\left\vert \frac{\partial\theta^{\prime}}{\partial\theta
}\right\vert \sqrt{g\left(  \theta^{\prime}\right)  }\text{, }d^{n}%
\theta\overset{\theta\rightarrow\theta^{\prime}}{\rightarrow}\left\vert
\frac{\partial\theta}{\partial\theta^{\prime}}\right\vert d^{n}\theta^{\prime
}\text{,} \label{pre1}%
\end{equation}
respectively. Hence, it follows that,%
\begin{equation}
\sqrt{g\left(  \theta\right)  }d^{n}\theta\overset{\theta\rightarrow
\theta^{\prime}}{\rightarrow}\sqrt{g\left(  \theta^{\prime}\right)  }%
d^{n}\theta^{\prime}\text{.} \label{pre3}%
\end{equation}
For further details on these issues, we refer the reader to Caticha's $2012$
tutorial \cite{tutorial}.

We have finally succeeded in obtaining all the elements necessary to express
the Riemannian volume element completely in terms of the coordinates $\theta$
and the known metric $g_{\mu\nu}\left(  \theta\right)  $, namely
$d\mathcal{V}_{\mathcal{M}}=\sqrt{g\left(  \theta\right)  }d^{n}\theta$. By
use of Fubini's theorem, the volume of any extended region on the manifold can
be expressed as follows\textbf{,}%
\begin{equation}
\mathcal{V}_{\mathcal{M}}\left(  s\right)  \overset{\text{def}}{=}%
\int_{\mathcal{D}_{\theta}^{\text{(geodesic)}}}d\mathcal{V}_{\mathcal{M}}%
=\int_{\mathcal{D}_{\theta}^{\text{(geodesic)}}}\sqrt{g\left(  \theta\right)
}d^{n}\theta\mathbf{=}\int d\theta^{1}\int d\theta^{2}\text{...}\int
\sqrt{g\left(  \theta^{1}\text{,..., }\theta^{n}\right)  }d\theta^{n}\text{,}
\label{due}%
\end{equation}
where $s\in%
\mathbb{R}
$. We remark that any permutation of the order of integration may be
considered in Eq. (\ref{due}) leaving the final result unchanged\textbf{. }The
integration space $\mathcal{D}_{\theta}^{\text{(geodesic)}}$ in Eq.
(\ref{due}) is defined as%
\begin{equation}
\mathcal{D}_{\theta}^{\text{(geodesic)}}\overset{\text{def}}{=}\left\{
\theta^{\kappa}\left(  \alpha\right)  :\theta^{\kappa}\left(  s_{0}\right)
\leq\theta^{\kappa}\leq\theta^{\kappa}\left(  s_{0}+s\right)  \right\}
\text{,} \label{is}%
\end{equation}
where $\kappa=1$,..., $n$ and $s_{0}\leq\alpha\leq s_{0}+s$ such that
$\theta^{\kappa}=\theta^{\kappa}\left(  \alpha\right)  $ satisfies Eq.
(\ref{geo1}). The integration space $\mathcal{D}_{\theta}^{\text{(geodesic)}}$
is an $n$-dimensional subspace of the whole (permitted) parameter space
$\mathcal{D}_{\theta}$. The elements of $\mathcal{D}_{\theta}%
^{\text{(geodesic)}}$ are the $n$-dimensional macrostates $\theta$ whose
components $\theta^{\kappa}$ are bounded by specified limits of integration
$\theta^{\kappa}\left(  s_{0}\right)  $ and $\theta^{\kappa}\left(
s_{0}+s\right)  $. The limits of integration are obtained via integration of
the $n$-dimensional set of geodesic equations. Now, by use of Eqs. (\ref{uno})
and (\ref{due}), we obtain%
\begin{equation}
\mathcal{V}_{\mathcal{M}}\left(  s\right)  \mathbf{=}\int d\theta^{1}\int
d\theta^{2}\text{...}\int\sqrt{g\left(  \theta^{1}\text{,..., }\theta
^{n}\right)  }d\theta^{n}=%
{\displaystyle\prod\limits_{\kappa=1}^{n}}
\int_{s_{0}}^{s_{0}+s}\sqrt{g_{\kappa}\left(  \theta^{\kappa}\left(
\alpha\right)  \right)  }\frac{d\theta^{\kappa}}{d\alpha}d\alpha\text{,}
\label{v-next}%
\end{equation}
where in Eq\textbf{.} (\ref{v-next}) we have made use of the following
equivalence,
\begin{equation}%
{\displaystyle\prod\limits_{\kappa=1}^{n}}
\left(  \int_{\theta^{\kappa}\left(  s_{0}\right)  }^{\theta^{\kappa}\left(
s_{0}+s\right)  }\sqrt{g_{\kappa}\left(  \theta^{\kappa}\right)  }%
d\theta^{\kappa}\right)  =%
{\displaystyle\prod\limits_{\kappa=1}^{n}}
\left(  \int_{s_{0}}^{s_{0}+s}\sqrt{g_{\kappa}\left(  \theta^{\kappa}\left(
\alpha\right)  \right)  }\frac{d\theta^{\kappa}}{d\alpha}d\alpha\right)
\text{.}%
\end{equation}
The extended volume defined in Eq. (\ref{v-next}) depends formally on both
$s_{0}$ (the selected initial instant from which we start computing a relevant
hyper-volume in $\mathcal{D}_{\theta}^{\text{(geodesic)}}$) and $s$
\textbf{(}the measure of the set of instances over which we observe the growth
or change of the hyper-volume). This procedure does not present a problem
since we are free to perform the additional two steps: 1) Take the limit for
$s_{0}$ approaching $0$. Indeed, we may also integrate in $ds_{0}$ with $0\leq
s_{0}\leq\epsilon$, where $\epsilon$ denotes the measure of the set of initial
instances. 2) Take the limit for $s$ approaching infinity (this is valid since
we are primarily interested in the asymptotic behavior of the IGE). One may be
concerned about the potential impact of selecting a different set of initial
conditions. Such issues however already enter at the level of the definition
of the functional forms for the geodesic trajectories $\theta^{\kappa}\left(
\alpha\right)  $ used to characterize the extended volume $\mathcal{V}%
_{\mathcal{M}}\left(  s\right)  $ over which one integrates. For this reason,
these issues do not affect the formal definition of the IGE.

\section{The Information Geometric Complexity}

Within the IGAC framework, we are interested in a probabilistic description of
the evolution of a given system in terms of its corresponding probability
distribution on $\mathcal{M}$ which is homeomorphic to $\mathcal{D}_{\theta}$.
For the sake of argument, consider the evolution of a system from
$s_{\text{initial}}$ to $s_{\text{final}}$. In the context of the present
probabilistic description of the MrE method \cite{caticha-giffin}, analysis of
this evolution is equivalent to studying the maximally probable path leading
from $\theta\left(  s_{\text{initial}}\right)  $ to $\theta\left(
s_{\text{final}}\right)  $. In order to quantify the complexity of such path,
we propose the so-called \emph{information geometric entropy} (IGE)
$\mathcal{S}_{\mathcal{M}}\left(  \tau\right)  $ as a good quantifier of
complexity \cite{carlo-PD, carlo-AMC}.

Within the context of our theoretical modeling scheme, the average dynamical
statistical volume\textbf{\ }$\mathcal{C}_{\mathcal{M}}\left(  \tau\right)
$\textbf{\ }[which we choose to name the \emph{information geometric
complexity} (IGC)] is defined as \cite{carlo-PD},%
\begin{equation}
\mathcal{C}_{\mathcal{M}}\left(  \tau\right)  \overset{\text{def}}{=}\frac
{1}{\tau}\int_{0}^{\tau}ds\mathcal{V}_{\mathcal{M}}\left(  s\right)  \text{.}
\label{rhs}%
\end{equation}
The IGC defined in Eq. (\ref{rhs}) represents the volume of the effective
parameter space explored by the system at affine time $\tau$. Alternatively,
$\mathcal{C}_{\mathcal{M}}\left(  \tau\right)  $ may be interpreted as the
temporal evolution of the system's uncertainty volume $\mathcal{C}%
_{\mathcal{M}}\left(  0\right)  $ after an affine temporal duration\textbf{\ }%
$\tau$ has elapsed. Its faithful geometric visualization may be highly non
trivial, especially in high-dimensional spaces.

The IGE, an indicator of temporal complexity of geodesic information flows, is
defined in terms of the IGC as follows,%
\begin{equation}
\mathcal{S}_{\mathcal{M}}\left(  \tau\right)  \overset{\text{def}}{=}%
\log\left[  \mathcal{C}_{\mathcal{M}}\left(  \tau\right)  \right]  \text{.}%
\end{equation}
The original idea underlying the formulation of the IGE was to provide a
quantitative means by which to encode dynamical information residing within
the hyper-volume $\mathcal{D}_{\theta}^{\text{(geodesic)}}$.

An analogue of our IGE\ can be found in the work of Myung, Balasubramanian and
Pitt \cite{minghia1, vijay, minghia2} as well as the work of Rodriguez
\cite{rodriguez04}. In both cases, the authors introduce a quantity that serve
to quantify intrinsic complexity, which in both cases is comprised of two
contributions. These two contributions are related to the notions of Bayesian
complexity penalty and minimum description length. In both cases, each of the
two contributions to intrinsic complexity are deemed to be inherent properties
of the statistical model describing the system under investigation since each
of the two contributions are independent of data. The first contribution is
defined as the product of the number of free parameters in the model (up to a
constant multiplicative factor) with the natural logarithm of the data sample
size (also up to a constant multiplicative factor). The second contribution to
the intrinsic complexity is defined as the logarithm of the integral of the
Riemannian volume element of a suitable parameter manifold of the model. The
Riemannian volume element is in turn specified in terms of the Fisher density
on the parameter manifold. With regard to this latter integral, the authors
\cite{minghia1} state: we will always cut off the ranges of the parameters to
ensure that volumes are finite. These ranges should be considered as part of
the functional form of the model. By construction, this volume is independent
of the parametrization. This second contribution to the geometric complexity
appearing in \cite{minghia1, rodriguez04} is reminiscent of our IGE both in
terms of its formal construction and its invariance under reparametrization of
the statistical model. In the case of our IGE however, there are two
noteworthy differences. Firstly, the\textbf{ }$\alpha$\textbf{-}dependent
lower and upper limits of integration (of the extended volume appearing in the
definition of the IGE) defines elements of\textbf{ }$\theta$\textbf{. }The
functional $\alpha$\textbf{-}dependence of these limits in turn depend upon
the nature of the geodesic equations underlying the statistical model being
considered. Indeed, the elements of $\theta$\textbf{ }are solutions of the
geodesic equations of the system. This system of geodesic equations is
integrated with suitable boundary conditions prior to the computation of the
hyper-volume\textbf{ }$\mathcal{D}_{\theta}^{\text{(geodesic)}}$. Secondly,
our IGE represents an affine temporal average of the\textbf{\ }$n$%
\textbf{-}fold integral of the Fisher density over maximum probability
trajectories (geodesics) and serves as a measure of the number of the
accessible macrostates in the statistical configuration manifold\textbf{
}$\mathcal{M}$\textbf{ }after a finite affine temporal increment\textbf{\ }%
$\tau$\textbf{. }The affine temporal average has been introduced in order to
average out the possibly very complex fine details of the entropic dynamical
description of the system on $\mathcal{M}$ \cite{caves}.

In what follows, we discuss the connection between the IGE and the
Kolmogorov-Sinai dynamical entropy. The notion of entropy is introduced, in
both classical and quantum physics, to quantify the missing information\ about
a system's coarse-grained state \cite{caves}. In the case of classical
systems, it is convenient to partition the phase space into fine-grained cells
of\ uniform volume $\Delta v$, labelled by an index $j$. In the absence of
knowledge of which cell the system occupies, one assigns probabilities $p_{j}%
$\ to each cell. In the limit of infinitesimal cells, and with the same state
of knowledge as in the previous coarse-grained case, one instead makes use of
the phase-space density $\rho\left(  X_{j}\right)  =\frac{p_{j}}{\Delta v}$.
Then, the\ asymptotic expression for the information required to characterize
a\textbf{ }particular coarse-grained trajectory up to time $t$\ is given by
the Shannon information entropy (measured in bits) \cite{caves},%
\begin{equation}
\mathcal{S}_{\text{classical}}^{\text{(chaotic)}}=-\int dX\rho\left(
X\right)  \log_{2}\left[  \rho\left(  X\right)  \Delta v\right]  =-\sum
_{j}p_{j}\log_{2}p_{j}\approx h_{\text{KS}}t\text{,} \label{KS}%
\end{equation}
where $\rho\left(  X\right)  $\ represents the phase-space density and
$p_{j}=\frac{v_{j}}{\Delta v}$\ is the probability of the corresponding
coarse-grained trajectory. The quantity $h_{\text{KS}}$ is the KS\ dynamical
entropy or metric entropy ($h_{\text{KS}}$ is actually an entropy rate, i.e.
an entropy per unit time), and represents the rate of information increase.
The quantity $S_{\text{classical}}^{\text{(chaotic)}}$\ represents the
missing\ information about which coarse-grained cell the system occupies.
According to the Alekseev-Brudno theorem \cite{alekseev}, the information
$I\left(  t\right)  $ (the quantity $I\left(  t\right)  $ is formally known as
the Kolmogorov algorithmic complexity \cite{kolmogorov-c}) associated with a
segment of a trajectory of length $\left\vert t\right\vert $ is asymptotically
equal to \cite{pesin}%
\begin{equation}
h_{\text{KS}}=\underset{\left\vert t\right\vert \rightarrow\infty}{\lim}%
\frac{I\left(  t\right)  }{\left\vert t\right\vert }. \label{KS entropy}%
\end{equation}
Stated in an alternative manner, the Alekseev-Brudno theorem implies that:
\emph{the KS entropy measures the algorithmic complexity of classical
trajectories} \cite{benattaman}. Within the IGAC, the information geometric
analogue $h_{\mathcal{M}}^{\text{KS}}$ of the KS dynamical entropy
$h_{\text{KS}}$ takes the form%
\begin{equation}
h_{\mathcal{M}}^{\text{KS}}=\lim_{\tau\rightarrow\infty}\left\{  \lim
_{\Delta\tau\rightarrow0}\left[  \frac{\mathcal{S}_{\mathcal{M}}\left(
\tau+\Delta\tau\right)  -\mathcal{S}_{\mathcal{M}}\left(  \tau\right)
}{\Delta\tau}\right]  \right\}  . \label{hks}%
\end{equation}

\section{Applications}

In the following, we outline ten selected applications concerning the
complexity characterization of geodesic paths on curved statistical manifolds
within the IGAC framework. Work featuring some of the early conceptual
developments of the IGAC\ framework can be found in \cite{PR9, PR7, PR6}. A
more updated overview of the IGAC appear in \cite{IG4, PR1, carlo-AMC}. The
initial series of applications presented in this Section represent systems of
arbitrary nature. Such systems are not only instructive, but also serve as
building blocks used to construct the more sophisticated, physically motivated
models appearing later in the section.

\subsection{Uncorrelated \textbf{Gaussian S\textbf{tatistical} Model}}

In \cite{carlo-PD, carlo-IJTP}, we apply the IGAC to study the dynamics of an
uncorrelated Gaussian statistical model specified by the probability
distribution%
\begin{equation}
p\left(  x|\theta\right)  =%
{\displaystyle\prod\limits_{k=1}^{l}}
p\left(  x_{k}|\mu_{k}\text{, }\sigma_{k}\right)  \text{, with }p\left(
x_{k}|\mu_{k}\text{, }\sigma_{k}\right)  \overset{\text{def}}{=}\frac{1}%
{\sqrt{2\pi\sigma_{k}^{2}}}\exp\left[  -\frac{\left(  x_{k}-\mu_{k}\right)
^{2}}{2\sigma_{k}^{2}}\right]  \text{,} \label{gauss}%
\end{equation}
where $x=\left(  x_{1}\text{,..., }x_{l}\right)  $ and $\theta=\left(  \mu
_{1}\text{,..., }\mu_{l}\text{, }\sigma_{1}\text{,..., }\sigma_{l}\right)  $.
The Gaussian distribution in Eq. (\ref{gauss})\ has $l$ degrees of freedom,
each one described by two pieces of relevant information, its mean expected
value $\mathbb{E}(x_{k})=\left\langle x_{k}\right\rangle =\mu_{k}$ and its
variance $\mathbb{E}(x_{k}-\mu_{k})^{2}=\Delta x_{k}=\sqrt{\left\langle
\left(  x_{k}-\left\langle x_{k}\right\rangle \right)  ^{2}\right\rangle
}=\sigma_{k}$ (Gaussian statistical macrostates). The line element
$ds^{2}=g_{\alpha\beta}\left(  \theta\right)  d\theta^{\alpha}d\theta^{\beta}$
($\alpha,\beta=1$,\ldots, $2l$) of the Fisher-Rao information metric
$g_{\alpha\beta}\left(  \theta\right)  $ on $\mathcal{M}$ is found to be
\cite{carlo-IJTP},%
\begin{equation}
ds^{2}=%
{\displaystyle\sum\limits_{k=1}^{l}}
\left(  \frac{1}{\sigma_{k}^{2}}d\mu_{k}^{2}+\frac{2}{\sigma_{k}^{2}}%
d\sigma_{k}^{2}\right)  \text{.}%
\end{equation}
This leads to consider a statistical model on a non-maximally symmetric
$2l$-dimensional statistical manifold $\mathcal{M}$. The IGE $\mathcal{S}%
_{\mathcal{M}}\left(  \tau\right)  $\ increases linearly in affine time and is
moreover, proportional to the number of degrees of freedom of the system,
\begin{equation}
\mathcal{S}_{\mathcal{M}}\left(  \tau\right)  \overset{\tau\rightarrow\infty
}{\approx}l\lambda_{\mathcal{M}}\tau\text{.}%
\end{equation}
The asymptotic linear growth of the IGE may be viewed as an
information-geometric analogue of the von Neumann entropy growth introduced by
Zurek-Paz \cite{Zurek}, a \textit{quantum} feature of chaos. The parameter
$\lambda_{\mathcal{M}}\in\mathbb{R}$ serves to characterize the family of
probability distributions on $\mathcal{M}$.

At this juncture, in anticipation of the correlated nature of the following
applications, we remark that in the presence of correlated constraints among
the microstates of a system, the\ product rule in Eq. (\ref{gauss}) assumes a
generalized form, while the metric tensor in Eq. (\ref{FRMetric}) will no
longer contain identically vanishing off-diagonal elements. Under such
scenarios the generalized\ version of the product rule in Eq. (\ref{gauss})
takes the form%
\begin{equation}
p_{\text{total}}\left(  x_{1},\ldots,x_{l}\right)  =%
{\displaystyle\prod\limits_{j=1}^{l}}
p_{j}\left(  x_{j}\right)  \overset{\text{correlations}}{\longrightarrow
}p_{\text{total}}^{\prime}\left(  x_{1},\ldots,x_{l}\right)  \neq%
{\displaystyle\prod\limits_{j=1}^{l}}
p_{j}\left(  x_{j}\right)  \text{,}%
\end{equation}
with%
\begin{equation}
p_{\text{total}}^{\prime}\left(  x_{1},\ldots,x_{l}\right)  =p_{l}\left(
x_{l}|x_{1},\ldots,x_{l-1}\right)  p_{l-1}\left(  x_{l-1}|x_{1},\ldots
,x_{l-2}\right)  \cdots p_{2}\left(  x_{2}|x_{1}\right)  p_{1}\left(
x_{1}\right)  \text{.}%
\end{equation}
On the one hand, correlations among the microvariables of a system may be
introduced via information-constraints of the form $x_{j}=f_{j}\left(
x_{1},\ldots,x_{j-1}\right)  ,$ $\forall j=2,\ldots,l$. In such a case%
\begin{equation}
p_{\text{total}}^{\prime}\left(  x_{1},\ldots,x_{l}\right)  =\delta\left(
x_{l}-f_{l}\left(  x_{1},\ldots,x_{l-1}\right)  \right)  \delta\left(
x_{l-1}-f_{l-1}\left(  x_{1},\ldots,x_{l-2}\right)  \right)  \cdots
\delta\left(  x_{2}-f_{2}\left(  x_{1}\right)  \right)  p_{1}\left(
x_{1}\right)  \text{,}%
\end{equation}
where the $j$-th probability distribution is given by%
\begin{equation}
p_{j}\left(  x_{j}\right)  =\int\cdots\int dx_{1}\cdots dx_{j-1}dx_{j+1}\cdots
dx_{l}p_{\text{total}}^{\prime}\left(  x_{1},\ldots,x_{l}\right)  \text{.}%
\end{equation}
On the other hand, correlations among the microvariables of a system may also
be introduced by means of the so-called correlation coefficients $\rho_{ij}$
\cite{roz},
\begin{equation}
\rho_{ij}=\rho\left(  x_{i},x_{j}\right)  \overset{\text{def}}{=}%
\frac{\left\langle x_{i}x_{j}\right\rangle -\left\langle x_{i}\right\rangle
\left\langle x_{j}\right\rangle }{\sigma_{i}\sigma_{j}}, \label{corr-coeff1}%
\end{equation}
with $\rho_{ij}\in\left(  -1,1\right)  $ and $i,j=1,\ldots,l$. The probability
distribution describing a $2l$-dimensional Gaussian model with non-vanishing
correlations is given by
\begin{equation}
p\left(  x|\theta\right)  =\frac{1}{\left[  \left(  2\pi\right)  ^{l}\det
C\left(  \theta\right)  \right]  ^{\frac{1}{2}}}\exp\left[  -\frac{1}%
{2}\left(  x-m\right)  ^{t}\cdot C^{-1}\left(  \theta\right)  \cdot\left(
x-m\right)  \right]  \neq%
{\displaystyle\prod\limits_{j=1}^{l}}
\left(  2\pi\sigma_{j}^{2}\right)  ^{-\frac{1}{2}}\exp\left[  -\frac{\left(
x_{j}-\mu_{j}\right)  ^{2}}{2\sigma_{j}^{2}}\right]  , \label{CG}%
\end{equation}
where $x=\left(  x_{1},\ldots,x_{l}\right)  $, $m=\left(  \mu_{1},\ldots
,\mu_{l}\right)  $ and $C\left(  \theta\right)  $ is the $\left(
2l\times2l\right)  $-dimensional (non-singular) covariance matrix. In the
following subsections we consider several statistical models with correlated microstates.

\subsection{Correlated Bivariate Gaussian Statistical Model}

In this application we consider a correlated bivariate Gaussian model
\cite{CNM-entropy}. The ratio between the IGC in presence and absence of
micro-correlations is explicitly computed, leading to an intriguing though not
yet deeply understood connection with the phenomenon of geometric frustration
\cite{SM99}. Specifically, we study in \cite{CNM-entropy} a $2D$ Gaussian
model specified by the probability distribution
\begin{equation}
p(x_{1},x_{2}|\mu,\sigma)=\frac{\exp\left\{  -\frac{1}{2\sigma^{2}(1-\rho
^{2})}\left[  (x_{1}-\mu)^{2}-2\rho(x_{1}-\mu)(x_{2}-\mu)+(x_{2}-\mu
)^{2}\right]  \right\}  }{2\pi\sigma^{2}\sqrt{1-\rho^{2}}}\text{,}%
\end{equation}
where the correlation coefficients $\rho_{ij}$ are defined in
(\ref{corr-coeff1}). The line element of the Fisher-Rao information metric
associated with $p(x_{1},x_{2}|\mu,\sigma)$ is given by%
\begin{equation}
ds^{2}=\frac{1}{\sigma^{2}\left(  1-\rho^{2}\right)  }d\mu^{2}+\frac{4}%
{\sigma^{2}}d\sigma^{2}\text{.}%
\end{equation}
The asymptotic expression of the IGC in this case is found to be%
\begin{equation}
\mathcal{C}(\tau)\overset{\tau\rightarrow\infty}{\approx}\left(  \frac
{4\sqrt{2}}{\sigma_{0}A_{1}}\right)  \frac{\sqrt{1+\rho}}{\tau}\text{ with
}\rho\in(-1,1)\text{,} \label{IGC2}%
\end{equation}
where $A_{1}\in\mathbb{R}$ is an integration constant, $\sigma_{0}=\left.
\sigma\left(  \tau\right)  \right\vert _{\tau=0}\in\mathbb{R}$ and
$\sigma\left(  \tau\right)  $\textbf{ }satisfies the geodesic equation\textbf{
(}\ref{geo1}). We may compare the ratio of the asymptotic expression of the
ICGs in the presence and absence of correlations, yielding the IGC ratio%
\begin{equation}
R_{\text{bivariate}}^{\text{strong}}(\rho)\overset{\text{def}}{=}%
\frac{\mathcal{C}_{\mathcal{M}}\left(  \tau\right)  }{\mathcal{C}%
_{\mathcal{M}}\left.  \left(  \tau\right)  \right\vert _{\rho=0}}=\sqrt
{1+\rho}\text{,} \label{ratio2}%
\end{equation}
where \textit{strong}\ represents the case in which the underlying microstates
of the system\textbf{ }are maximally connected. The ratio $R_{\text{bivariate}%
}^{\text{strong}}(\rho)$ results in an increasing \emph{monotone} function of
$\rho$. From Eq\textbf{.} (\ref{ratio2}), it is evident that for
anti-correlated variables, an increase in one variable gives rise to a
corresponding decrease in the remaining variable; this result implies that
variables become more distant and therefore more distinguishable relative to
the Fisher-Rao information metric. By contrast, for positively correlated
variables, an increase or decrease in one variable always predicts a similarly
directed change in the remaining variable. In this scenario, variables do not
become more distant and consequently, are not more distinguishable relative to
the Fisher-Rao metric. This result seems to suggest that when
anti-correlations are present, evolution on the statistical manifold induced
by the system reduces in complexity.

\subsection{Correlated Trivariate Gaussian Statistical Model}

In \cite{CNM-entropy}, we study the IG of a trivariate Gaussian statistical
model where the multivariate normal joint distributions for $n$ real-valued
microstates $x_{1},\ldots,x_{n}$ is given by
\begin{equation}
p(x|\theta)=\frac{1}{\sqrt{(2\pi)^{n}\det C}}\exp\left[  -\frac{1}{2}%
(x-\mu)^{t}\cdot{C}^{-1}\cdot(x-\mu)\right]  , \label{PxT}%
\end{equation}
with $C$ denoting the $n\times n$ symmetric, positive definite covariance
matrix with entries $c_{ij}=\mathbb{E}(x_{i}x_{j})-\mathbb{E}(x_{i}%
)\mathbb{E}(x_{j})$, ${i,j=1,\ldots,n}$. It is assumed that the mean and
variance of each of the three microstates are (i.e. $\mu_{x}=\mu_{y}=\mu
_{z}=\mu$ and $\sigma_{x}=\sigma_{y}=\sigma_{z}=\sigma$). Furthermore, each
model has different correlational structure between microstates. In this
section we consider a Gaussian statistical model in Eq\textbf{.} (\ref{PxT})
for the case $n=3$. The covariance matrices corresponding to these cases are
given by \cite{FMP},
\begin{equation}
C_{1}=\sigma^{2}\left(
\begin{array}
[c]{ccc}%
1 & \rho & 0\\
\rho & 1 & 0\\
0 & 0 & 1
\end{array}
\right)  \text{, }C_{2}=\sigma^{2}\left(
\begin{array}
[c]{ccc}%
1 & \rho & \rho\\
\rho & 1 & 0\\
\rho & 0 & 1
\end{array}
\right)  \text{, and }C_{3}=\sigma^{2}\left(
\begin{array}
[c]{ccc}%
1 & \rho & \rho\\
\rho & 1 & \rho\\
\rho & \rho & 1
\end{array}
\right)  \text{.}%
\end{equation}

\subsubsection{Case 1}

First, we consider the trivariate Gaussian statistical model corresponding to
the case $C=C_{1}$. The line element of the Fisher-Rao information metric
corresponding to this choice of covariance matrix $C_{1}$ is given by,%
\begin{equation}
ds^{2}=\frac{3+\rho}{(1+\rho)\sigma^{2}}d\mu^{2}+\frac{6}{\sigma^{2}}%
d\sigma^{2}\text{.}%
\end{equation}
The asymptotic expression of the IGC in this case is given by
\begin{equation}
\mathcal{C}_{\mathcal{M}}(\tau)\overset{\tau\rightarrow\infty}{\approx}\left(
\frac{6\sqrt{6}}{\sigma_{0}A_{1}}\right)  \sqrt{\frac{1+\rho}{3+\rho}}%
\ \frac{1}{\tau}\text{ with }\rho\in(-1,1)\text{,} \label{IGC31}%
\end{equation}
where $A_{1}\in\mathbb{R}$ is an integration constant, $\sigma_{0}=\left.
\sigma\left(  \tau\right)  \right\vert _{\tau=0}\in\mathbb{R}$ and\textbf{
}$\sigma\left(  \tau\right)  $\textbf{ }satisfies the geodesic
equation\textbf{ (}\ref{geo1}). Comparing Eq. (\ref{IGC31}) in the presence
and absence of correlations yields the IGC ratio%
\begin{equation}
R_{\text{trivariate}}^{\text{weak}}(\rho)\overset{\text{def}}{=}%
\frac{\mathcal{C}_{\mathcal{M}}(\tau)}{\mathcal{C}_{\mathcal{M}}\left.
(\tau)\right\vert _{\rho=0}}=\sqrt{3}\sqrt{\frac{1+\rho}{3+\rho}}\text{,}
\label{ratio31}%
\end{equation}
where \textit{weak}\ represents the case in which the underlying microstates
of the system are minimally connected. Observe that $R_{\text{bivariate}%
}^{\text{weak}}(\rho)$ is an increasing monotone function of the argument
$\rho\in(-1,1)$.

\subsubsection{Case 2}

In the second case, we consider the trivariate Gaussian statistical model
corresponding to Eq. (\ref{PxT}) with the choice $C=C_{2}$. For this choice of
covariance matrix, the condition $C>0$ constrains the correlation coefficient
to the range $\rho\in(-\frac{\sqrt{2}}{2},\frac{\sqrt{2}}{2})$. The\textbf{
}Fisher-Rao information metric line element associated with this model is
given by%
\begin{equation}
ds^{2}=\frac{3-4\rho}{(1-2\rho^{2})\sigma^{2}}d\mu^{2}+\frac{6}{\sigma^{2}%
}d\sigma^{2}\text{.}%
\end{equation}
The asymptotic behavior of the IGC is found to be
\begin{equation}
\mathcal{C}_{\mathcal{M}}(\tau)\overset{\tau\rightarrow\infty}{\approx}\left(
\frac{6\sqrt{6}}{\sigma_{0}A_{1}}\right)  \sqrt{\frac{1-2\rho^{2}}{3-4\rho}%
}\ \frac{1}{\tau}. \label{IGC32}%
\end{equation}
Then, by means of comparison of\ Eq.\textbf{ }(\ref{IGC32}) in the presence
and absence of correlations yield the IGC ratio
\begin{equation}
R_{\text{trivariate}}^{\text{mildly weak}}(\rho)\overset{\text{def}}{=}%
\frac{\mathcal{C}_{\mathcal{M}}(\tau)}{\mathcal{C}_{\mathcal{M}}\left.
(\tau)\right\vert _{\rho=0}}=\sqrt{3}\sqrt{\frac{1-2\rho^{2}}{3-4\rho}},
\label{ratio32}%
\end{equation}
where \textit{mildly weak}\ represents the case in which the underlying
microstates of the system\textbf{ }are neither minimally nor maximally
connected. The ratio $R_{\text{trivariate}}^{\text{mildly weak}}(\rho)$ is a
function of the argument $\rho\in(-\frac{\sqrt{2}}{2},\frac{\sqrt{2}}{2})$ and
it attains the maximal value of $\sqrt{\frac{3}{2}}$ at $\rho=\frac{1}{2}$,
while in the extrema of the interval $(-\frac{\sqrt{2}}{2},\frac{\sqrt{2}}%
{2})$ it tends to zero.

\subsubsection{Case 3}

As our final case for this example, we consider the trivariate Gaussian
statistical model of Eq. (\ref{PxT}) when $C=C_{3}$. In this case, the
condition $C>0$ requires that the correlation coefficient assume values in the
range $\rho\in(-\frac{1}{2},1)$. The Fisher-Rao information metric line
element associated with this model is given by%
\begin{equation}
ds^{2}=\frac{3}{(1+2\rho)\sigma^{2}}d\mu^{2}+\frac{6}{\sigma^{2}}d\sigma
^{2}\text{.}%
\end{equation}
The asymptotic behavior of the IGC reduces to%
\begin{equation}
\mathcal{C}_{\mathcal{M}}(\tau)\overset{\tau\rightarrow\infty}{\approx}\left(
\frac{12}{\sigma_{0}A_{1}}\right)  \frac{\sqrt{1+2\rho}}{\tau}\text{,}
\label{IGC33}%
\end{equation}
where $A_{1}\in\mathbb{R}$ is an integration constant, $\sigma_{0}=\left.
\sigma\left(  \tau\right)  \right\vert _{\tau=0}\in\mathbb{R}$ and\textbf{
}$\sigma\left(  \tau\right)  $ satisfies the geodesic equation\textbf{
(}\ref{geo1}). The comparison of Eq\textbf{.} (\ref{IGC33}) in the presence
and absence of correlations yield the IGC ratio
\begin{equation}
R_{\text{trivariate}}^{\text{strong}}(\rho)\overset{\text{def}}{=}%
\frac{\mathcal{C}_{\mathcal{M}}(\tau)}{\mathcal{C}_{\mathcal{M}}\left.
(\tau)\right\vert _{\rho=0}}=\sqrt{1+2\rho}\text{,} \label{ratio3}%
\end{equation}
where\textit{ strong}\ represents a maximally connected lattice underlying the
trivariate microstates of the system. It is obvious that the ratio
$R_{\text{trivariate}}^{\text{strong}}(\rho)$ is an increasing monotonic
function of the argument $\rho\in(-\frac{1}{2},1)$. Observe that the growth of
$R_{\text{trivariate}}^{\text{mildly weak}}(\rho)$ terminates at the critical
value of $\rho_{\text{peak}}=\frac{1}{2}$ where $R_{\text{trivariate}%
}^{\text{mildly weak}}(\rho_{\text{peak}})=R_{\text{trivariate}}%
^{\text{strong}}(\rho_{\text{peak}})$. Interestingly, these conclusions are
quite similar to those presented for the bivariate case. There is however, a
key-feature of the IGC worth emphasizing when transitioning from the
two-dimensional to the three-dimensional manifolds associated with cases
exhibiting maximal connectedness among the microvariables of the system.\ In
particular, the effects of negative and positive correlations are both
\emph{amplified} relative to the respective scenarios lacking correlations,
such that
\begin{equation}
\frac{R_{\text{trivariate}}^{\text{strong}}(\rho)}{R_{\text{bivariate}%
}^{\text{strong}}(\rho)}=\sqrt{\frac{1+2\rho}{1+\rho}}\text{,}
\label{ratio3su2}%
\end{equation}
where $\rho\in(-\frac{1}{2},1)$. The above results enables us to conclude that
the implementation of entropic inferences on higher-dimensional manifolds in
the presence of anti-correlations $\left[  \text{i.e. }\rho\in\left(
-\frac{1}{2},0\right)  \right]  $ is less complex than that on
lower-dimensional manifolds as is evident form Eq. (\ref{ratio3su2}). The
converse is true in the presence of positive-correlations $\left[  \text{i.e.
}\rho\in\left(  0,1\right)  \right]  $.

\subsection{Complexity Reduction Arising from Microcorrelations}

In \cite{carloPA2010}, we consider a three-dimensional Gaussian model
specified by the probability distribution%
\begin{equation}
p_{\text{correlated}}(x,y|\mu_{x},\mu_{y},\sigma)=\frac{\exp\left\{  -\frac
{1}{2\sigma^{2}(1-\rho^{2})}\left[  (x-\mu_{x})^{2}-2\rho(x-\mu_{x})(y-\mu
_{y})+(y-\mu_{y})^{2}\right]  \right\}  }{2\pi\sigma^{2}\sqrt{1-\rho^{2}}}%
\end{equation}
where $\sigma\in\left(  0,\infty\right)  $, $\mu_{x}$ and $\mu_{y}\in\left(
-\infty,\infty\right)  $ and $\rho\in\left(  0,+1\right)  $, from which the
Fisher-Rao information metric line element%
\begin{equation}
ds_{\text{correlated}}^{2}=\frac{1}{\sigma^{2}}\left[  \frac{1}{1-\rho^{2}%
}\left(  d\mu_{x}^{2}+d\mu_{x}^{2}+2\rho d\mu_{x}d\mu_{y}\right)
+4d^{2}\sigma\right]
\end{equation}
is obtained. The asymptotic expression of the IGC is given by%
\begin{equation}
\mathcal{C}_{\text{correlated}}(\tau,\rho)\overset{\tau\rightarrow\infty
}{\approx}\frac{a^{2}}{2\sigma_{0}\mathcal{A}^{\frac{3}{2}}}\sqrt
{\frac{4\left(  4-\rho^{2}\right)  }{\left(  2-2\rho^{2}\right)  ^{2}}}%
\frac{1}{\tau}%
\end{equation}
where $\sigma_{0}=\left.  \sigma\left(  \tau\right)  \right\vert _{\tau=0}$,
$\mathcal{A}\overset{\text{def}}{=}\frac{A_{1}^{2}+A_{2}^{2}-\rho A_{1}A_{2}%
}{4\left(  1-\rho^{2}\right)  }$ and without loss of generality, $A_{1}%
=-A_{2}=a\in%
\mathbb{R}
$. Upon comparison of the asymptotic expressions of the IGCs in presence and
absence of microcorrelations, where $\mathcal{C}_{\text{uncorrelated}}%
(\tau,0)=\mathcal{C}_{\text{correlated}}(\tau,\rho\rightarrow0)$, we obtain%
\begin{equation}
\mathcal{C}_{\text{correlated}}(\tau,\rho)\overset{\tau\rightarrow\infty
}{\approx}\mathcal{F}\left(  \rho\right)  \cdot\mathcal{C}%
_{\text{uncorrelated}}(\tau,0)\text{,} \label{temp9}%
\end{equation}
where $0\leq\mathcal{F}\left(  \rho\right)  \leq1$ is defined as,%
\begin{equation}
\mathcal{F}\left(  \rho\right)  \overset{\text{def}}{=}\frac{1}{2^{\frac{5}%
{2}}}\sqrt{\frac{4\left(  4-\rho^{2}\right)  }{\left(  2-2\rho^{2}\right)
^{2}}}\left(  \frac{2+\rho}{4\left(  1-\rho^{2}\right)  }\right)  ^{-\frac
{3}{2}}. \label{Funct}%
\end{equation}
The quantity $\mathcal{F}\left(  \rho\right)  $ is a monotonic decreasing
function for any value of the correlation coefficient $\rho$ in the open
interval $\left(  0,+1\right)  $. In essence, it represents an asymptotic
power law decay of the IGC at a rate determined by $\rho$. The result of this
analysis, captured in (\ref{temp9}), suggest that systems containing
microcorrelations experience an asymptotic compression of the explored
statistical microstates at a faster rate than in the absence of
microcorrelations. This finding represents an explicit connection between the
behavior of the experimentally observable macroscopic quantities of a
statistical system on the information encoded in the correlational structure
underlying the system's microscopic degrees of freedom.

\subsection{Complexity Reduction Arising from Macrocorrelations}

In \cite{CM, cafaro-mancini}, we consider a $4l$-dimensional Gaussian
statistical model specified by the probability distribution%
\begin{equation}
p\left(  x|\theta\right)  =%
{\displaystyle\prod\limits_{k=1}^{2l}}
p\left(  x_{k}|\mu_{k}\text{, }\sigma_{k}\right)  \text{, }p\left(  x_{k}%
|\mu_{k}\text{, }\sigma_{k}\right)  \overset{\text{def}}{=}\frac{1}{\sqrt
{2\pi\sigma_{k}^{2}}}\exp\left[  -\frac{\left(  x_{k}-\mu_{k}\right)  ^{2}%
}{2\sigma_{k}^{2}}\right]  \text{,}%
\end{equation}
with $x\equiv\left(  x_{1}\text{,..., }x_{2l}\right)  $ and $\theta
\equiv\left(  \mu_{1}\text{,..., }\mu_{2l}\text{, }\sigma_{1}\text{,...,
}\sigma_{2l}\right)  $, from which the information line element
\begin{equation}
ds^{2}=%
{\displaystyle\sum\limits_{j=1}^{2l}}
\frac{1}{\sigma_{j}^{2}}\left(  d\mu_{j}^{2}+2d\sigma_{j}^{2}\right)  ,
\end{equation}
is obtained. By subjecting the statistical microstates $x_{k}$ to a set of
$2l$ embedding constraints,%
\begin{equation}
\sigma_{2j}=\sigma_{2j-1}\text{ and, }\mu_{2j}=\mu_{2j}\left(  \mu
_{2j-1}\text{, }\sigma_{2j-1}\right)  \text{ with }j=1,...,l,
\label{correlational1}%
\end{equation}
the probability distribution $p\left(  x|\theta\right)  $ reduces to the
$2l$-dimensional embedded Gaussian statistical model%
\begin{equation}
p_{\text{embedded}}\left(  x|\theta\right)  =%
{\displaystyle\prod\limits_{j=1}^{l}}
p\left(  x_{2j-1}\text{, }x_{2j}|\mu_{2j-1}\text{, }\sigma_{2j-1}\right)
\text{,} \label{scenario2}%
\end{equation}
with $x=\left(  x_{1}\text{,..., }x_{2l}\right)  $ and $\theta=\left(  \mu
_{1}\text{, }\mu_{3}\text{,..., }\mu_{2l-1}\text{; }\sigma_{1}\text{, }%
\sigma_{3}\text{,..., }\sigma_{2l-1}\right)  $ where $p\left(  x_{2j-1}\text{,
}x_{2j}|\mu_{2j-1}\text{, }\sigma_{2j-1}\right)  $ is defined as%
\begin{equation}
p\left(  x_{2j-1}\text{, }x_{2j}|\mu_{2j-1}\text{, }\sigma_{2j-1}\right)
\overset{\text{def}}{=}\frac{1}{2\pi\sigma_{2j-1}^{2}}\exp\left[
-\frac{\left(  x_{2j-1}-\mu_{2j-1}\right)  ^{2}+\left[  x_{2j}-\mu_{2j}\left(
\mu_{2j-1}\text{, }\sigma_{2j-1}\right)  \right]  ^{2}}{2\sigma_{2j-1}^{2}%
}\right]  \text{,}%
\end{equation}
and $j=1$,..., $l$. The model in Eq. (\ref{scenario2}) leads to the
information line element%
\begin{equation}
ds_{\text{embedded}}^{2}=%
{\displaystyle\sum\limits_{j=1}^{l}}
\frac{1}{\sigma_{2j-1}^{2}}\left(  d\mu_{2j-1}^{2}+2\rho_{2j-1}d\mu
_{2j-1}d\sigma_{2j-1}+2d\sigma_{2j-1}^{2}\right)  \text{,} \label{ntfm}%
\end{equation}
with the coefficients $\rho_{2j-1}$ defined as%
\begin{equation}
\rho_{2j-1}\overset{\text{def}}{=}\frac{\frac{\partial\mu_{2j}}{\partial
\mu_{2j-1}}\frac{\partial\mu_{2j}}{\partial\sigma_{2j-1}}}{\left[  1+\left(
\frac{\partial\mu_{2j}}{\partial\mu_{2j-1}}\right)  ^{2}\right]  ^{\frac{1}%
{2}}\left[  2+\frac{1}{2}\left(  \frac{\partial\mu_{2j}}{\partial\sigma
_{2j-1}}\right)  ^{2}\right]  ^{\frac{1}{2}}}, \label{rk}%
\end{equation}
where the explicit expressions of such coefficients depend on the functional
parametric form given to the embedding constraints in Eq\textbf{.}
(\ref{correlational1}). From\ Eq\textbf{. }(\ref{rk}) it follows that the
coefficients $\rho_{2j-1}$ are non-zero if and only if $\mu_{2j}$ depends on
both $\mu_{2j-1}$ and $\sigma_{2j-1}$. Therefore, we conclude that the
emergence of non-vanishing off-diagonal terms in Eq\textbf{.} (\ref{ntfm})
arise due to the presence of a correlation among the statistical variables on
the larger manifold and are therefore characterized by the macroscopic
correlation coefficients $\rho_{2j-1}$. Motivated by these considerations, we
will name the coefficients $\rho_{2j-1}$ \emph{macroscopic correlation
coefficients}. The IGE was determined to have the form\textbf{ }\cite{CM,
cafaro-mancini}
\begin{equation}
\mathcal{S}_{\mathcal{M}}\left(  \tau;l,\lambda_{k},\rho_{k}\right)
\overset{\tau\rightarrow\infty}{\approx}\log\left[  \Lambda_{1}\left(
r_{k}\right)  +\frac{\Lambda_{2}\left(  \rho_{k},\lambda_{k}\right)  }{\tau
}\right]  ^{l}\text{,} \label{hodor}%
\end{equation}
provided $\rho_{k}=\rho_{s}$ $\forall k$ and $s=1,\ldots,l$, with
\begin{equation}
\Lambda_{1}\left(  \rho_{k}\right)  \overset{\text{def}}{=}\frac{2\rho
_{k}\sqrt{2-\rho_{k}^{2}}}{1+\sqrt{\Delta\left(  \rho_{k}\right)  }}\text{,
}\Lambda_{2}\left(  \rho_{k},\lambda_{k}\right)  \overset{\text{def}}{=}%
\frac{\sqrt{\Delta\left(  \rho_{k}\right)  \left(  2-\rho_{k}^{2}\right)
}\log\left[  \Sigma\left(  \rho_{k},\lambda_{k},\alpha_{\pm}\right)  \right]
}{\rho_{k}\lambda_{k}}\text{, and }\alpha_{\pm}\left(  \rho_{k}\right)
\overset{\text{def}}{=}\frac{1}{2}\left(  3\pm\sqrt{\Delta\left(  \rho
_{k}\right)  }\right)  \text{.}%
\end{equation}
The quantity%
\begin{equation}
\Sigma\left(  \rho_{k},\lambda_{k},\alpha_{\pm}\right)  \overset{\text{def}%
}{=}-\frac{\Xi_{k}}{4\lambda_{k}}\frac{1+\sqrt{\Delta\left(  \rho_{k}\right)
}}{1-\sqrt{\Delta\left(  \rho_{k}\right)  }}\sqrt{\frac{2\alpha_{-}\left(
\rho_{k}\right)  }{\alpha_{+}\left(  \rho_{k}\right)  }}>0\text{, }\forall
\rho\in\lbrack0,1)
\end{equation}
is a strictly positive function of its arguments, where $\Xi_{k}$ and
$\lambda_{k}$ are \emph{real}, \emph{positive} constants of integration and%
\begin{equation}
\Delta\left(  \rho_{k}\right)  \overset{\text{def}}{=}1+4\rho_{k}^{2}\text{.}%
\end{equation}
It is evident from Eq. (\ref{hodor}) that the IGE is characterized by a power
law decay, whereby the power is specified by the cardinality of the
microscopic degrees of freedom associated with correlated macroscopic
information. Furthermore, the IGE attains a maximal value quantified by the
set $\left\{  \rho_{k}\right\}  $. The relevance of these finding is twofold:
first, it provides a compact description of the effect of microscopic
information on (experimentally observable) macroscopic variables; second, it
provides quantitative evidence that the information geometric complexity of a
system decreases in the presence of correlational structures.

\subsection{Suppression of Classical Chaos from Quantum-like constraints: The
Uncorrelated Case}

Building upon the results obtained in \cite{PR3, PR4, PR2}, we investigate in
\cite{OSID} a $3Du$ (three-dimensional and uncorrelated) Gaussian statistical
model specified by the probability distribution%
\begin{equation}
p_{3Du}\left(  x\text{, }y|\mu_{x}\text{, }\sigma_{x}\text{, }\sigma
_{y}\right)  \overset{\text{def}}{=}\frac{1}{2\pi\sigma_{x}\sigma_{y}}%
\exp\left[  -\frac{1}{2\sigma_{x}^{2}}\left(  x-\mu_{x}\right)  ^{2}-\frac
{1}{2\sigma_{y}^{2}}y^{2}\right]  ,
\end{equation}
whose Fisher-Rao information metric line element is given by%
\begin{equation}
ds_{3Du}^{2}=\frac{1}{\sigma_{x}^{2}}\left(  d\mu_{x}^{2}+2d\sigma_{x}%
^{2}\right)  +\frac{2}{\sigma_{y}^{2}}d\sigma_{y}^{2}.
\end{equation}
We then compare our analysis to that of a $2Du$ (two-dimensional and
uncorrelated)\ Gaussian statistical model obtained from the higher-dimensional
model $p_{3Du}\left(  x\text{, }y|\mu_{x}\text{, }\sigma_{x}\text{, }%
\sigma_{y}\right)  $ via the introduction of a macroscopic information
constraint%
\begin{equation}
\sigma_{x}\sigma_{y}=\Sigma^{2}\text{, }\Sigma^{2}\in%
\mathbb{R}
_{0}^{+} \label{MC}%
\end{equation}
that resembles the quantum mechanical canonical minimum uncertainty relation,
which leads to the $2Du$ statistical model%
\begin{equation}
p_{2Du}\left(  x\text{, }y|\mu_{x}\text{, }\sigma\right)  \overset{\text{def}%
}{=}\frac{1}{2\pi\Sigma^{2}}\exp\left[  -\frac{1}{2\sigma^{2}}\left(
x-\mu_{x}\right)  ^{2}-\frac{\sigma^{2}}{2\Sigma^{4}}y^{2}\right]  \text{,}
\label{GP}%
\end{equation}
where\textbf{\ }$x$\textbf{\ }denotes the position of a particle and\textbf{
}$y$\textbf{\ }its conjugate momentum\textbf{.} The line element of the
Fisher-Rao information metric associated with $p_{2Du}\left(  x\text{, }%
y|\mu_{x}\text{, }\sigma\right)  $ is given by,
\begin{equation}
ds_{2Du}^{2}=\frac{1}{\sigma^{2}}\left(  d\mu_{x}^{2}+4d\sigma^{2}\right)
\text{.}%
\end{equation}
It was determined that for the $3Du$ model, the IGE takes the form%
\begin{equation}
\mathcal{S}_{\mathcal{M}}^{\text{(}3Du\text{)}}\left(  \tau\right)
\overset{\tau\rightarrow\infty}{\approx}\lambda_{+}^{\prime}\tau\text{,}
\label{IGE2}%
\end{equation}
where $\lambda_{+}^{\prime}\in%
\mathbb{R}
^{+}$. In the $2Du$ case, it was found that%
\begin{equation}
\mathcal{S}_{\mathcal{M}}^{\text{(}2Du\text{)}}\left(  \tau\right)
\overset{\tau\rightarrow\infty}{\approx}\lambda_{+}\tau\text{,} \label{IGE1}%
\end{equation}
where $\lambda_{+}=$ $\frac{\lambda_{+}^{\prime}}{\sqrt{2}}$ $\in%
\mathbb{R}
^{+}$. By comparing $\mathcal{S}_{\mathcal{M}}^{\text{(}3Du\text{)}}$ with
$\mathcal{S}_{\mathcal{M}}^{\text{(}2Du\text{)}}$, one observes%
\begin{equation}
\mathcal{S}_{\mathcal{M}}^{\text{(}2Du\text{)}}\left(  \tau\right)
\overset{\tau\rightarrow\infty}{\approx}\left[  \left(  \frac{\lambda_{+}%
}{\lambda_{+}^{\prime}}\right)  \cdot\mathcal{S}_{\mathcal{M}}^{\text{(}%
3Du\text{)}}\left(  \tau\right)  \right]  \text{ with }\frac{\lambda_{+}%
}{\lambda_{+}^{\prime}}=\frac{1}{\sqrt{2}}<1\text{.} \label{IGEfinal}%
\end{equation}

\subsection{Suppression of Classical Chaos from Quantum-like constraints: The
Correlated Case}

In \cite{SoftChaos}, we study a correlated $3Dc$ \textbf{(}three-dimensional
and correlated\textbf{) }Gaussian statistical model with uncorrelated
microstates specified by the probability distribution%
\begin{equation}
p_{3Dc}\left(  x\text{, }y|\mu_{x}\text{, }\sigma_{x}\text{, }\sigma_{y}%
;\rho\right)  =\dfrac{1}{2\pi\sigma_{x}\sigma_{y}\sqrt{1-\rho^{2}}}\exp\left[
\frac{-1}{2\left(  1-\rho^{2}\right)  }\left(  \frac{\left(  x-\mu_{x}\right)
^{2}}{\sigma_{x}^{2}}+\frac{y^{2}}{\sigma_{y}^{2}}-\frac{2\rho\left(
x-\mu_{x}\right)  y}{\sigma_{x}\sigma_{y}}\right)  \right]
\end{equation}
whose line element of the Fisher-Rao information metric associated with
$p_{3Dc}\left(  x\text{, }y|\mu_{x}\text{, }\sigma_{x}\text{, }\sigma
_{y}\right)  $ is given by%
\begin{equation}
ds_{3Dc}^{2}=\frac{1}{1-\rho^{2}}\frac{d\mu_{x}^{2}}{\sigma_{x}^{2}}%
+\frac{2-\rho^{2}}{1-\rho^{2}}\frac{d\sigma_{x}^{2}}{\sigma_{x}^{2}}%
+\frac{2-\rho^{2}}{1-\rho^{2}}\frac{d\sigma_{y}^{2}}{\sigma_{y}^{2}}%
-\frac{2\rho^{2}}{1-\rho^{2}}\frac{d\sigma_{x}d\sigma_{y}}{\sigma_{x}%
\sigma_{y}}\text{.}%
\end{equation}
We then compare our analysis to that of a $2Dc$ \textbf{(}two-dimensional and
correlated\textbf{) }Gaussian statistical model obtained from the
higher-dimensional model $p_{3Dc}\left(  x\text{, }y|\mu_{x}\text{, }%
\sigma_{x}\text{, }\sigma_{y}\right)  $ via introduction of a
\emph{covariance} constraint%
\begin{equation}
\sigma_{xy}=\rho\sigma_{x}\sigma_{y}\text{,} \label{CovC}%
\end{equation}
where the parameter $\rho$ is the correlation coefficient between $x$ and $y$
and assumes values within the ranges\textbf{\ }$-1\leq\rho\leq1$. Applying the
macroscopic constraint in Eq\textbf{.} (\ref{MC}) to the covariance constraint
in Eq\textbf{.} (\ref{CovC}) yields the combined constraint
\begin{equation}
\sigma_{xy}=\rho\Sigma^{2}\text{ with }\sigma_{x}\sigma_{y}=\Sigma^{2}\in%
\mathbb{R}
_{0}^{+}%
\end{equation}
which when applied to $p_{3Dc}\left(  x\text{, }y|\mu_{x}\text{, }\sigma
_{x}\text{, }\sigma_{y};\rho\right)  $ leads to%
\begin{equation}
p_{2Dc}\left(  x\text{, }y|\mu_{x}\text{, }\sigma;\rho\right)  \overset
{\text{def}}{=}\dfrac{1}{2\pi\Sigma^{2}\sqrt{1-\rho^{2}}}\exp\left\{
\frac{-1}{2\left(  1-\rho^{2}\right)  }\left[  \frac{\left(  x-\mu_{x}\right)
^{2}}{\sigma^{2}}+\frac{y^{2}\sigma^{2}}{\Sigma^{4}}-\frac{2\rho\left(
x-\mu_{x}\right)  y}{\Sigma^{2}}\right]  \right\}  \text{.}%
\end{equation}
The line element of the Fisher-Rao information metric associated with
$p_{2Dc}\left(  x\text{, }y|\mu_{x}\text{, }\sigma\right)  $ is given by
\begin{equation}
ds_{2Dc}^{2}=\frac{1}{\sigma^{2}\left(  1-\rho^{2}\right)  }d\mu_{x}^{2}%
+\frac{4}{\sigma^{2}\left(  1-\rho^{2}\right)  }d\sigma^{2}\text{.}%
\end{equation}
It was determined that for the correlated $2Dc$ model, the IGE takes the form%
\begin{equation}
\mathcal{S}_{\mathcal{M}}^{\text{(}2Dc\text{)}}\left(  \tau\right)
=\log\left[  \mathcal{C}_{\mathcal{M}}^{\text{(}2Du\text{)}}\left(
\tau\right)  \right]  \overset{\tau\rightarrow\infty}{\approx}\sigma
_{0}\lambda_{+}\tau\text{,}%
\end{equation}
where the correlated IGC is given by
\begin{equation}
\mathcal{C}_{\mathcal{M}}^{\text{(}2Dc\text{)}}\left(  \tau\right)
\overset{\tau\rightarrow\infty}{\approx}\frac{1}{\left(  1-\rho^{2}\right)
}\mathcal{C}_{\mathcal{M}}^{\text{(}2Du\text{)}}\left(  \tau\right)  \text{
with }\mathcal{C}_{\mathcal{M}}^{\text{(}2Du\text{)}}\left(  \tau\right)
\overset{\tau\rightarrow\infty}{\approx}\left[  \left(  \frac{\mu_{0}%
+2\sigma_{0}}{\sigma_{0}^{2}\lambda_{+}}\right)  \frac{\exp\left(  \sigma
_{0}\lambda_{+}\tau\right)  }{\tau}\right]  \text{,} \label{IGCfinale}%
\end{equation}
where $\lambda_{+}\in%
\mathbb{R}
^{+}$, $\sigma_{0}\overset{\text{def}}{=}\sigma\left(  \tau=0\right)  $ and
$\mu_{0}\overset{\text{def}}{=}\mu\left(  \tau=0\right)  $. By comparing
$\mathcal{S}_{\mathcal{M}}^{\text{(}2Du\text{)}}$ with $\mathcal{S}%
_{\mathcal{M}}^{\text{(}2Dc\text{)}}$, one observes%
\begin{equation}
\mathcal{S}_{\mathcal{M}}^{\text{(}2Dc\text{)}}\overset{\tau\rightarrow\infty
}{\approx}\mathcal{S}_{\mathcal{M}}^{\text{(}2Du\text{)}}\text{.}%
\end{equation}
The IGE does not change asymptotically for either the correlated or
uncorrelated $2D$ models considered above. Equation (\ref{IGCfinale})
quantitatively demonstrates that the IGC $\mathcal{C}_{\mathcal{M}}%
^{\text{(}2Dc\text{)}}$diverges as the correlation coefficient - introduced
via the constraint in Eq\textbf{.} (\ref{CovC}) - approaches unity. As
expected, the two cases are identical for $\rho=0$.

\subsection{Random Frequency Macroscopic Anisotropic Inverted Harmonic
Oscillators}

Building upon the results obtained in \cite{caticha-cafaro}, we present in
\cite{carlo-CSF, EJTP} an information geometric analogue of the Zurek-Paz
quantum chaos criterion in the \emph{classical reversible limit}. This analogy
is illustrated by applying our modeling scheme to a set of\textbf{\ }%
$l$\textbf{-}uncoupled, three-dimensional anisotropic, inverted harmonic
oscillators (IHOs) characterized by a Ohmic distributed frequency spectrum. In
this application, we consider a manifold whose metric line element\textbf{ }is
given by\textbf{ }\cite{carlo-CSF, EJTP}%
\begin{equation}
ds^{2}=\left[  1-\Phi\left(  \theta\right)  \right]  \delta_{ab}d\theta
^{a}d\theta^{b}\text{ with }\Phi\left(  \theta\right)  =\overset{l}%
{\underset{k=1}{\sum}}u_{k}\left(  \theta^{k}\right)  \text{, }%
\end{equation}
where $\delta_{ab}$ is the identity matrix of dimension $l$ and
\begin{equation}
u_{k}\left(  \theta^{k}\right)  =-\frac{1}{2}\omega_{k}^{2}\left(  \theta
^{k}\right)  ^{2}\text{, with }\theta^{k}=\theta^{k}\left(  s\right)  \text{.}%
\end{equation}
Upon making a suitable change of the affine parameter featured in the geodesic
equations from $s$ to $\tau$ in such a manner that $ds^{2}=2\left(
1-\Phi\right)  ^{2}d\tau^{2}$, we can obtain a simplified form for these
differential equations describing a set of macroscopic inverted harmonic
oscillators (IHOs). Since the $l$-Newtonian equations of motion for each IHO
is given by%
\begin{equation}
\frac{d^{2}\theta^{j}}{d\tau^{2}}-\omega_{j}^{2}\theta^{j}=0,\text{ }\forall
j=1,\ldots,l\text{,}%
\end{equation}
the asymptotic behavior of such macrostates on manifold $\mathcal{M}%
_{\text{IHO}}^{\left(  l\right)  }$ is determined to be%
\begin{equation}
\theta^{j}\left(  \tau\right)  \overset{\tau\rightarrow\infty}{\approx}\Xi
_{j}e^{\omega_{j}\tau}\text{, with }\Xi_{j}\in%
\mathbb{R}
,\text{ }\forall j=1,\ldots,l\text{.}%
\end{equation}
Thus, after some analysis, the IGC and IGE were found to have the forms%
\begin{equation}
\mathcal{C}_{\mathcal{M}_{\text{IHO}}^{\left(  l\right)  }}\left(  \tau
;\omega_{1},\ldots,\omega_{l}\right)  \overset{\tau\rightarrow\infty\text{ }%
}{\approx}\frac{1}{l}\frac{1}{2^{\frac{l}{2}}}\Xi^{2l}\left(  \frac{\xi
^{2}\Omega^{2}}{2}\right)  ^{\frac{l}{2}}\frac{\exp\left(  \frac{l}{2}%
\xi\Omega\tau\right)  }{\tau}\text{, } \label{inter6}%
\end{equation}
and%
\begin{equation}
\mathcal{S}_{\mathcal{M}_{\text{IHO}}^{\left(  l\right)  }}\left(  \tau
;\omega_{1},\ldots,\omega_{l}\right)  \overset{\tau\rightarrow\infty}{\approx
}\Omega\tau\text{,} \label{Fin}%
\end{equation}
respectively, where
\begin{equation}
\Omega=\overset{l}{\underset{i=1}{\sum}}\omega_{i},\text{ }\Xi_{j}\in%
\mathbb{R}
,\forall j=1,\ldots,l
\end{equation}
and in Eq. (\ref{inter6}) we assumed that $\Xi_{i}=\Xi_{j}\equiv\Xi$ $\forall
i,$ $j=1,\ldots,l$ with\textbf{ }$\xi$\textbf{ }being a positive real constant
that depends on the specific nature of the system being considered
\cite{carlo-CSF, EJTP}. Note that it was further assumed that $l\rightarrow
\infty$ such that the spectrum of frequencies transitions into a continuum
with linearly distributed spectrum (Ohmic frequency spectrum),
\begin{equation}
\theta_{\text{Ohmic}}\left(  \omega\right)  =\frac{2\omega}{\Omega
_{\text{cut-off}}^{2}}\text{, where }\underset{0}{\overset{\Omega
_{\text{cut-off}}}{\int}}\theta_{\text{Ohmic}}\left(  \omega\right)
d\omega=1\text{ and }\Omega_{\text{cut-off}}=\gamma\Omega\text{ with }%
\gamma\in%
\mathbb{R}
\text{.}%
\end{equation}
Equation (\ref{Fin}) displays an asymptotic, linear IGE growth for the
generalized set of inverted harmonic oscillators and serves to extend the
result of Zurek-Paz to an arbitrary ensemble of anisotropic inverted harmonic
oscillators \cite{zurek1} within a classical IG setting. This example may be
viewed as the IG counterpart of the Zurek-Paz model used to investigate the
effects of decoherence in quantum chaos. In their work, Zurek and Paz
considered a single unstable harmonic oscillator characterized by a potential
\begin{equation}
V\left(  x\right)  =-\frac{\Omega^{2}x^{2}}{2} \label{vimo}%
\end{equation}
coupled to an external environment. Note that the quantity $\Omega$ in Eq.
(\ref{vimo}) represents the Lyapunov exponent. In the reversible classical
limit \cite{zurek2}, the von Neumann entropy of such a system increases
linearly at a rate determined by the Lyapunov exponent according to,
\begin{equation}
\mathcal{S}_{\text{quantum}}^{\text{(chaotic)}}\left(  \tau\right)
\overset{\tau\rightarrow\infty}{\approx}\Omega\tau\text{.} \label{hud1}%
\end{equation}
Equation (\ref{Fin}) is effectively the classical IG analog of Eq. (\ref{hud1}).

\subsection{Regular and Chaotic Quantum Spin Chains}

In \cite{carloMPLB, cafaroPA}, we proposed an IG characterization of
integrable and chaotic energy level statistics of a quantum anti-ferromagnetic
Ising spin chain. In this example, we encode relevant information about the
spin-chain in a suitable composite probability distribution taking account of
both the quantum spin chain and the configuration of the external magnetic
field in which the spins are immersed. Specifically, for the integrable case
an anti-ferromagnetic Ising chain is immersed in a transverse, homogeneous
magnetic field
\begin{equation}
\vec{B}_{\text{transverse}}=B_{\perp}\hat{B}_{_{\perp}}\text{,}%
\end{equation}
which has only \emph{one} component $B_{\perp}$, where the level spacing
distribution of its spectrum is Poisson distributed. It is known from
information theory that the Exponential distribution is identified as the
maximum entropy distribution if only \emph{one} piece of information (in this
case the expectation value of the transverse magnetic field) is known\textbf{.
}For this reason\textbf{, }we consider a Poisson distribution
\begin{equation}
p_{A}^{\text{(Poisson)}}\left(  x_{A}|\mu_{A}\right)  =\frac{1}{\mu_{A}}%
\exp\left(  -\frac{x_{A}}{\mu_{A}}\right)  ,
\end{equation}
coupled to an Exponential bath
\begin{equation}
p_{B}^{\text{(Exponential)}}\left(  x_{B}|\mu_{B}\right)  =\frac{1}{\mu_{B}%
}\exp\left(  -\frac{x_{B}}{\mu_{B}}\right)  ,
\end{equation}
which together, gives rise to the composite statistical model%
\begin{equation}
P^{\text{(integrable)}}\left(  x_{A},x_{B}|\mu_{A},\mu_{B}\right)  =\frac
{1}{\mu_{A}\mu_{B}}\exp\left[  -\left(  \frac{x_{A}}{\mu_{A}}+\frac{x_{B}}%
{\mu_{B}}\right)  \right]  ,
\end{equation}
inducing manifold $\mathcal{M}^{\text{(integrable)}}$. The microstate $x_{A}$
represents the spacing of the energy levels while the macrostate $\mu_{A}$ is
the average level spacing; the microstate $x_{B}=\left\vert -\vec{\mu}%
\cdot\vec{B}\right\vert =\left\vert -\mu B\cos\varphi\right\vert $\ is
identified with the intensity of the magnetic field, $\varphi$\ is the tilt
angle and the macrostate $\mu_{B\text{ }}$is the average transverse magnetic
field intensity. In the transverse case, $\varphi=0$\ and therefore
$x_{B}=B\equiv B_{\perp}$. This model represents our best guess as justified
by the observation that the magnitude of the magnetic field is a relevant
quantity in this scenario. The components of the transverse magnetic field are
varied during the transition from integrable to chaotic regimes. In the
integrable regime, the magnetic field intensity is set to the well-defined
value $\left\langle x_{B}\right\rangle =\mu_{B}$.

In the chaotic case, an antiferromagnetic Ising chain is immersed in a
\emph{tilted}, homogeneous magnetic field
\begin{equation}
\vec{B}_{\text{tilted}}=B_{\perp}\hat{B}_{\perp}+B_{\parallel}\hat
{B}_{\parallel},
\end{equation}
comprised of \emph{two} components\textbf{ }$B_{\perp}$ and $B_{\parallel}$,
with the level spacing distribution of its spectrum given by the Wigner-Dyson
distribution of Poisson form. It is known from information theory that the
Gaussian distribution is identified as the maximum entropy distribution when
only two pieces of information are known (in this case the expectation value
and the variance). For these reasons, we consider a Wigner-Dyson distribution
\begin{equation}
p_{A^{\prime}}^{\text{(Wigner-Dyson)}}\left(  x_{A}^{\prime}|\mu_{A}^{\prime
}\right)  =\frac{\pi x_{A}^{\prime}}{2\mu_{A}^{\prime2}}\exp\left(  -\frac{\pi
x_{A}^{\prime2}}{4\mu_{A}^{\prime2}}\right)  ,
\end{equation}
coupled to a Gaussian bath
\begin{equation}
p_{B^{\prime}}^{\text{(Gaussian)}}\left(  x_{B}^{\prime}|\mu_{B}^{\prime
},\sigma_{B}^{\prime}\right)  =\frac{1}{\sqrt{2\pi\sigma_{B}^{\prime2}}}%
\exp\left(  -\frac{\left(  x_{B}^{\prime}-\mu_{B}^{\prime}\right)  ^{2}%
}{2\sigma_{B}^{\prime2}}\right)  ,
\end{equation}
which together, gives rise to the composite statistical model%
\begin{equation}
P^{\text{(chaotic)}}\left(  x_{A}^{\prime},x_{B}^{\prime}|\mu_{A}^{\prime}%
,\mu_{B}^{\prime},\sigma_{B}^{\prime}\right)  =\frac{\pi\left(  2\pi\sigma
_{B}^{\prime2}\right)  ^{-\frac{1}{2}}}{2\mu_{A}^{\prime2}}x_{A}^{\prime}%
\exp\left[  -\left(  \frac{\pi x_{A}^{\prime2}}{4\mu_{A}^{\prime2}}%
+\frac{\left(  x_{B}^{\prime}-\mu_{B}^{\prime}\right)  ^{2}}{2\sigma
_{B}^{\prime2}}\right)  \right]  , \label{p-exp}%
\end{equation}
inducing manifold $\mathcal{M}^{\text{(chaotic)}}$. Note that $B_{\perp}$ and
$B_{\parallel}$\ are transverse and longitudinal magnetic field intensities,
respectively. The microstate $x_{B}^{\prime}$ is identified with the intensity
of the tilted magnetic field, while the macrostate $\mu_{B\text{ }}^{\prime}%
$is the average intensity of the magnetic energy arising from the interaction
of the tilted magnetic field with the magnetic moment of the spin $\frac{1}%
{2}$\ particle and $\sigma_{B}^{\prime}$ is its covariance.\ During the
transition from the integrable to chaotic regimes, the magnetic field is being
experimentally varied. Specifically, the magnetic field is being tilted while
its two components $B_{\perp}$ and $B_{\parallel}$ are simultaneously being
varied. Our best guess based upon knowledge of the experimental mechanism that
drives the transitions between the two regimes is that the the microstate $\mu
B\cos\varphi$ is Gaussian-distributed during this change. In the chaotic
regime, the magnetic field intensity is set to a well-defined value
$\left\langle x_{B}^{\prime}\right\rangle =\mu_{B}^{\prime}$\ with covariance
$\sigma_{B}^{\prime}=\sqrt{\left\langle \left(  x_{B}^{\prime}-\left\langle
x_{B}^{\prime}\right\rangle \right)  ^{2}\right\rangle }$.

The line element $ds_{\text{integrable}}^{2}$ of the Fisher-Rao information
metric on $\mathcal{M}^{\text{(integrable)}}$ is given by%
\begin{equation}
ds_{\text{integrable}}^{2}=ds_{\text{Poisson}}^{2}+ds_{\text{Exponential}}%
^{2}=\frac{1}{\mu_{A}^{2}}d\mu_{A}^{2}+\frac{1}{\mu_{B}^{2}}d\mu_{B}^{2}.
\label{regular case}%
\end{equation}
Applying the IGAC to the line element in Eq. (\ref{regular case}) leads to
polynomial growth in $\mathcal{C}_{\mathcal{M}}^{\text{(integrable)}}$ and
logarithmic IGE growth \cite{cafaroPA, carloMPLB}, according to
\begin{equation}
\mathcal{C}_{\mathcal{M}}^{\text{(integrable)}}\left(  \tau\right)
\overset{\tau\rightarrow\infty}{\approx}\exp(c_{\text{IG}}^{\prime}%
)\tau^{c_{\text{IG}}},\text{ }\mathcal{S}_{\mathcal{M}}^{\text{(integrable)}%
}\left(  \tau\right)  \overset{\tau\rightarrow\infty}{\approx}c_{\text{IG}%
}\log\tau+c_{\text{IG}}^{\prime}. \label{polinomio}%
\end{equation}
The quantity\textbf{\ }$c_{\text{IG}}$\textbf{\ }is a constant that is
proportional to the number of Exponential probability distributions within the
composite distribution utilized in the computation of\textbf{ }the IGE;
$c_{\text{IG}}^{\prime}$\ is a constant that depends on the values assumed by
the statistical macrostates\textbf{\ }$\mu_{A}$\textbf{\ }and\textbf{\ }%
$\mu_{B}$. Equations (\ref{polinomio}) may be interpreted as the IG analogues
of the computational complexity and the entanglement entropy\ defined in
standard quantum information theory, respectively.

The Fisher-Rao information metric line element $ds_{\text{chaotic}}^{2}$ on
$\mathcal{M}^{\text{(chaotic)}}$ is given by%
\begin{equation}
ds_{\text{chaotic}}^{2}=ds_{\text{Wigner-Dyson}}^{2}+ds_{\text{Gaussian}}%
^{2}=\frac{4}{\mu_{A}^{\prime2}}d\mu_{A}^{\prime2}+\frac{1}{\sigma_{B}%
^{\prime2}}d\mu_{B}^{\prime2}+\frac{2}{\sigma_{B}^{\prime2}}d\sigma
_{B}^{\prime2}. \label{chaotic case}%
\end{equation}
Applying the IGAC machinery to the line element in Eq. (\ref{chaotic case}),
we obtain exponential growth for $\mathcal{C}_{\mathcal{M}}^{\text{(chaotic)}%
}$ and linear IGE growth \cite{cafaroPA, carloMPLB},
\begin{equation}
\mathcal{C}_{\mathcal{M}}^{\text{(chaotic)}}\left(  \tau\right)  \overset
{\tau\rightarrow\infty}{\approx}q_{\text{IG}}\exp\left(  K_{\text{IG}}%
\tau\right)  ,\text{ and }\mathcal{S}_{\mathcal{M}}^{\text{(chaotic)}}\left(
\tau\right)  \overset{\tau\rightarrow\infty}{\approx}K_{\text{IG}}\tau\text{,}
\label{exponential}%
\end{equation}
respectively. The constant\ $q_{\text{IG}}$\ acts to encode information
concerning the initial conditions of the macrostates parametrizing
$\mathcal{M}^{\text{(chaotic)}}$. The constant $K_{\text{IG}}$ given by,%
\begin{equation}
K_{\text{IG}}\overset{\tau\rightarrow\infty}{\approx}\frac{d\mathcal{S}%
_{\mathcal{M}}\left(  \tau\right)  }{d\tau}%
\end{equation}
is the model parameter of the chaotic system and depends on the temporal
evolution of the macrostates of the system. As in the integrable case,
equations (\ref{exponential}) may be interpreted as the IG analogues of the
computational complexity and the entanglement entropy defined in standard
quantum information theory, respectively.

\subsection{Scattering Induced Quantum Entanglement}

Building upon the results obtained in \cite{PR5}, we implement in \cite{kim2,
kim1} a hybrid approach (standard quantum theory combined with IG techniques)
to the modeling of scattering induced quantum entanglement \cite{epr,
Schrodinger}. In particular, we performed an IG analysis of two identical,
distinguishable Continuous Variable Quantum Systems (CVQS) with Gaussian
distributed continuous degrees of freedom (i.e. spinless, non-relativistic
point particles of mass $m$, each represented by minimum uncertainty Gaussian
wave packets where both particles\ are initially located far from each other,
a linear distance\ $R_{\mathrm{o}}$, and each being characterized before
collision by initial average momentum\ $\left\langle \mathbf{p}_{1}%
\right\rangle _{\mathrm{o}}=p_{\mathrm{o}}$\ and\ $\left\langle \mathbf{p}%
_{2}\right\rangle _{\mathrm{o}}=-p_{\mathrm{o}}$,\ respectively,\ with equal
momentum dispersion\ $\sigma_{\mathrm{o}}$) that are prepared independently.
The two wave packets interact via a scattering process\ mediated by an
interaction ($s$-wave scattering) potential $V(x)$ and separate again. Note
that by CVQS we refer to quantum mechanical systems on which one can, in
principle, perform measurements of certain observables whose eigenvalue
spectrum is continuous. For such a system,\ a complete set of commuting
observables is furnished by the momentum operators of each particle
\cite{Harshman1, Harshman2}. The interaction potential\ $V\left(  x\right)
$\ is isotropic and is active over a short range\ $L$\ such that $V\left(
x\right)  =V$ for $0\leq x\leq L$ and\ $V\left(  x\right)  \approx
0$\ for\ $x>L$ where $V$ denotes the height (for $V>0$; repulsive potential)
or depth (for $V<0$; attractive potential) of the potential.

We investigate the quantum entanglement quantified in terms of a scalar
quantity, the purity\textbf{ }$\mathcal{P}\overset{\text{def}}{=}%
\mathrm{Tr}\left(  \rho_{A}^{2}\right)  $ of the two-particle state function
describing the system generated by such a scattering event. We note that
$\mathrm{Tr}$ denotes the standard quantum-mechanical trace operation on the
operator\textbf{ }$\rho_{A}^{2}$. In the case being considered, $\rho
_{A}\overset{\text{def}}{=}\mathrm{Tr}_{B}\left(  \rho_{AB}\right)  $ is the
reduced density matrix that describes particle $A$ and\textbf{ }$\rho_{AB}$
represents the two-particle density matrix associated with the
post-collisional two-particle wave function. Note that $\mathrm{Tr}_{B}$
denotes the partial trace over the particle $B$. The operation of computing
$\mathrm{Tr}_{B}$ is usually referred to as tracing-out system $B$. Briefly
speaking, the reduced density operator $\mathrm{Tr}_{B}\left(  \rho
_{AB}\right)  $ is the correct tool to use when analyzing physical properties
that belong solely to $A$ \cite{mosca}. For pure bipartite\textbf{ }states,
the smaller the value of $\mathcal{P}$, the higher the entanglement. Thus, the
loss of purity furnishes an indicator of the degree of entanglement, where a
disentangled product state corresponds to $\mathcal{P}=1$.

\subsubsection{The Pre and Post Collision Scenarios}

The normalized, separable, two-particle Gaussian wave function representing
the situation prior to collision is prescribed by \cite{Wang, Law}%
\begin{equation}
\psi^{\text{(Pre)}}\left(  k_{1},k_{2}\right)  =\psi_{1}\left(  k_{1}\right)
\otimes\psi_{2}\left(  k_{2}\right)  \text{, with }\psi_{1\left(  2\right)
}\left(  k_{1\left(  2\right)  }\right)  =a\left(  k_{1\left(  2\right)
},\left\langle k_{1\left(  2\right)  }\right\rangle _{\mathrm{o}}%
;\sigma_{k\mathrm{o}}\right)  e^{i\left(  k_{1\left(  2\right)  }-\left\langle
k_{1\left(  2\right)  }\right\rangle _{\mathrm{o}}\right)  q_{1\left(
2\right)  }}, \label{eq:int1}%
\end{equation}
where
\begin{equation}
a\left(  k_{1\left(  2\right)  },\left\langle k_{1\left(  2\right)
}\right\rangle _{\mathrm{o}};\sigma_{k\mathrm{o}}\right)  \overset{\text{def}%
}{=}\left(  \frac{1}{2\pi\sigma_{k\mathrm{o}}^{2}}\right)  ^{1/4}\exp\left[
-\frac{\left(  k_{1\left(  2\right)  }-\left\langle k_{1\left(  2\right)
}\right\rangle _{\mathrm{o}}\right)  ^{2}}{4\sigma_{k\mathrm{o}}^{2}}\right]
, \label{int2}%
\end{equation}
and $k_{1\left(  2\right)  }=\frac{p_{1\left(  2\right)  }}{\hbar}\in\left(
-\infty,+\infty\right)  $, $\left\langle k_{1\left(  2\right)  }\right\rangle
_{\mathrm{o}}=\frac{\left\langle p_{1\left(  2\right)  }\right\rangle
_{\mathrm{o}}}{\hbar}=\pm\frac{p_{\mathrm{o}}}{\hbar}=\pm k_{\mathrm{o}}$,
$\sigma_{k\mathrm{o}}=\frac{\sigma_{\mathrm{o}}}{\hbar}$, $q_{1\left(
2\right)  }=\mp\frac{1}{2}R_{\mathrm{o}}$, and\textbf{ }$\hbar$\textbf{ }is
the reduced Planck constant. The pre-collision probability density\textbf{\ }%
$\left\vert \psi^{\text{(Pre)}}\left(  k_{1},k_{2}\right)  \right\vert ^{2}$
takes the form%
\begin{equation}
p_{\text{\textrm{QM}}}^{\text{(Pre)}}\overset{\text{def}}{=}\left\vert
\psi^{\text{(Pre)}}\left(  k_{1},k_{2},t\right)  \right\vert ^{2}=\frac
{1}{2\pi\sigma_{k\mathrm{o}}^{2}}\exp\left[  -\frac{\left(  k_{1}-\left\langle
k_{1}\right\rangle _{\mathrm{o}}\right)  ^{2}+\left(  k_{2}-\left\langle
k_{2}\right\rangle _{\mathrm{o}}\right)  ^{2}}{2\sigma_{k\mathrm{o}}^{2}%
}\right]  .
\end{equation}
After collision, the wave function for the two-particle system in the long
time limit takes the form \cite{Wang}%
\begin{equation}
\psi^{\text{(Post)}}\left(  k_{1},k_{2},t\right)  =\left(  N\right)
^{-1/2}\left[  \psi_{1}\left(  k_{1}\right)  \psi_{2}\left(  k_{2}\right)
e^{-i\hbar\left(  k_{1}^{2}+k_{2}^{2}\right)  t/\left(  2m\right)
}+\varepsilon\psi_{\mathrm{scat}}\left(  k_{1},k_{2},t\right)  \right]  ,
\label{psi0}%
\end{equation}
where $N$ and $\varepsilon$ are normalization constants and the
single-particle wave function $\psi_{1\left(  2\right)  }\left(  k_{1\left(
2\right)  }\right)  $ is specified via Eq\textbf{.} (\ref{eq:int1}). The
quantity $\psi^{\text{(Post)}}\left(  k_{1},k_{2},t\right)  $ can be rewritten
as
\begin{align}
\psi^{\text{(Post)}}\left(  k_{1},k_{2},t\right)   &  =\left(  N\right)
^{-1/2}\left(  \frac{1}{2\pi\sigma_{k\mathrm{o}}^{2}}\right)  ^{1/2}%
\exp\left[  -\frac{K^{2}+4\left(  k-k_{\mathrm{o}}\right)  ^{2}}%
{8\sigma_{k\mathrm{o}}^{2}}\right] \nonumber\\
&  \times\left[  1+\varrho\left(  k\right)  \right]  e^{-i\left(
k-k_{\mathrm{o}}\right)  R_{\mathrm{o}}-i\hbar K^{2}t/\left(  2M\right)
-i\hbar k^{2}t/\left(  2\mu\right)  }, \label{psi1}%
\end{align}
where we adopt the one-dimensional center of mass and relative
coordinates,\ whose conjugate momenta are defined as $K\overset{\text{def}}%
{=}k_{1}+k_{2}\in\left(  -\infty,+\infty\right)  $, $k\overset{\text{def}}%
{=}\frac{1}{2}\left(  k_{1}-k_{2}\right)  \in\left(  -\infty,+\infty\right)
$, $M\overset{\text{def}}{=}2m$\textbf{\ }is the total mass,\textbf{\ }%
$\mu\overset{\text{def}}{=}\frac{m}{2}$\textbf{\ }is the reduced mass, and
$\varrho\left(  k\right)  $ is given by
\begin{equation}
\varrho\left(  k\right)  \overset{\text{def}}{=}\frac{4i\left(  k_{\mathrm{o}%
}-i\sigma_{k\mathrm{o}}^{2}R_{\mathrm{o}}\right)  k^{2}f\left(  k\right)
}{\sigma_{k\mathrm{o}}^{2}}, \label{sc6}%
\end{equation}
where $f\left(  k\right)  \overset{\text{def}}{=}\frac{e^{i2\theta\left(
k\right)  }-1}{2ik}$ is the $s$-wave scattering amplitude due to the $s$-wave
scattering phase shift $\theta\left(  k\right)  $ and $i=\sqrt{-1}$\textbf{
}is the standard imaginary unit. The post-collision probability
density\textbf{\ }$\left\vert \psi^{\text{(Post)}}\left(  k_{1},k_{2}%
,t\right)  \right\vert ^{2}$ takes the form
\begin{equation}
p_{\text{\textrm{QM}}}^{\text{(Post)}}\overset{\text{def}}{=}\left\vert
\psi^{\text{(Post)}}\left(  k_{1},k_{2},t\right)  \right\vert ^{2}\approx
\frac{1}{N}\frac{\exp\left\{  -\frac{1}{2\left(  1-\rho_{\mathrm{QM}}%
^{2}\right)  }\left[  \frac{\left(  k_{1}-k_{\mathrm{o}}\right)  ^{2}}%
{\sigma_{k\mathrm{o}}^{2}}-2\rho_{\mathrm{QM}}\frac{\left(  k_{1}%
-k_{\mathrm{o}}\right)  \left(  k_{2}+k_{\mathrm{o}}\right)  }{\sigma
_{k\mathrm{o}}^{2}}+\frac{\left(  k_{2}+k_{\mathrm{o}}\right)  ^{2}}%
{\sigma_{k\mathrm{o}}^{2}}\right]  \right\}  }{2\pi\sigma_{k\mathrm{o}}%
^{2}\sqrt{1-\rho_{\mathrm{QM}}^{2}}}, \label{eq:psi_sq2}%
\end{equation}
where the constant $N$ in Eq. (\ref{eq:psi_sq2}) has been determined so as to
normalize the integral. The quantity $\rho_{\mathrm{QM}}$ appearing in
(\ref{eq:psi_sq2}) is defined as
\begin{equation}
\rho_{\mathrm{QM}}\overset{\text{def}}{=}\sqrt{8\left(  2k_{\mathrm{o}}%
^{2}+\sigma_{k\mathrm{o}}^{2}\right)  R_{\mathrm{o}}a_{\mathrm{s}}}\ll1,
\label{rho12}%
\end{equation}
where the parameter $a_{\mathrm{s}}$ has the dimension of length and is
defined as the $s$-wave scattering length \cite{Landau}, which comes from
\begin{equation}
f\left(  k_{\mathrm{o}}\right)  =\frac{e^{i\theta_{\mathrm{o}}}\sin
\theta_{\mathrm{o}}}{k_{\mathrm{o}}}\overset{\theta\left(  k_{\mathrm{o}%
}\right)  \ll1}{\approx}\frac{\theta\left(  k_{\mathrm{o}}\right)
}{k_{\mathrm{o}}}+\mathcal{O}\left(  \theta^{2}\right)  ,
\end{equation}
where
\begin{equation}
\tan\theta_{\mathrm{o}}=\frac{k_{\mathrm{o}}\tan\left(  k_{r}L\right)
-k_{r}\tan\left(  k_{\mathrm{o}}L\right)  }{k_{r}+k_{\mathrm{o}}\tan\left(
k_{\mathrm{o}}L\right)  \tan\left(  k_{r}L\right)  }\overset{k_{\mathrm{o}%
}L=p_{\mathrm{o}}L/\hbar\ll1}{\approx}\theta_{\mathrm{o}}=\theta\left(
k_{\mathrm{o}}\right)  \approx-\frac{\rho\left(  k_{\mathrm{o}}L\right)  ^{3}%
}{3}%
\end{equation}
denotes the $s$-wave scattering phase shift, under the\textbf{\ }assumption of
low energy $s$-wave scattering with $k_{\mathrm{o}}L=p_{\mathrm{o}}L/\hbar
\ll1$. From our IG analysis,%
\begin{equation}
k_{r}\overset{\text{def}}{=}\sqrt{1-\rho}k_{\mathrm{o}}\text{, }0<x<L
\end{equation}
where%
\begin{align}
k_{r}  &  =\frac{\sqrt{2\mu\left(  \mathcal{E}-V\right)  }}{\hbar
},\;0<x<L\text{, and}\\
k_{\mathrm{o}}  &  =\frac{\sqrt{2\mu\mathcal{E}}}{\hbar},\;x>L. \label{eq:k_0}%
\end{align}
We remark that the quantity $\rho_{\mathrm{QM}}$\textbf{ }in Eq\textbf{.
}(\ref{rho12}) can be related to the two-particle squeezing\textbf{\ }%
parameters\textbf{\ }found in \cite{Serafini} for example.

\subsubsection{Information Geometric\ Modeling of\ Quantum\ Entanglement}

Our \emph{first conjecture} is that the quantum entanglement produced by a
head-on collision between two Gaussian wave packets are macroscopic
manifestations emerging from underlying microscopic statistical structures.
For this reason, we model the pre and post-collisional scenarios as limiting
cases of uncorrelated and correlated Gaussian statistical models%
\begin{equation}
p_{\text{\textrm{QM}}}^{\text{(Pre)}}\approx p_{\text{IG}}^{\text{(uncorr)}%
}\left(  x,y|\mu_{x},\mu_{y},\sigma\right)  =\frac{1}{2\pi\sigma^{2}}%
\exp\left[  -\frac{\left(  x-\mu_{x}\right)  ^{2}}{2\sigma^{2}}-\frac{\left(
y-\mu_{y}\right)  ^{2}}{2\sigma^{2}}\right]  , \label{eq:new_P}%
\end{equation}
and%
\begin{equation}
p_{\text{\textrm{QM}}}^{\text{(Post)}}\approx p_{\text{IG}}^{\text{(corr)}%
}\left(  x,y|\mu_{x},\mu_{y},\sigma;\rho\right)  =\frac{\exp\left\{  -\frac
{1}{2\left(  1-\rho^{2}\right)  }\left[  \frac{\left(  x-\mu_{x}\right)  ^{2}%
}{\sigma^{2}}-2\rho\frac{\left(  x-\mu_{x}\right)  \left(  y-\mu_{y}\right)
}{\sigma^{2}}+\frac{\left(  y-\mu_{y}\right)  ^{2}}{\sigma^{2}}\right]
\right\}  }{2\pi\sigma^{2}\sqrt{1-\rho^{2}}}%
\end{equation}
respectively, with the following identifications: $x\rightarrow k_{1}$,
$y\rightarrow k_{2}$; $\left\langle x\right\rangle =\mu_{x}\rightarrow
\mu_{k_{1}}\overset{\text{def}}{=}+k_{\mathrm{o}}$, $\left\langle
y\right\rangle =\mu_{y}\rightarrow\mu_{k_{2}}\overset{\text{def}}%
{=}-k_{\mathrm{o}}$ and $\sigma\rightarrow\sigma_{k_{\mathrm{o}}}$. The
correlation coefficient $\rho$ is defined in Eq. (\ref{corr-coeff1}). In the
present example, the correlation coefficient $\rho$ is considered to have
compact support over the line segment $[0,1)$, and is identified with the
quantum entanglement strength $\rho_{\mathrm{QM}}$, such that $\rho
=\rho_{\mathrm{QM}}$. In order
to\ furnish\ an\ IG\ interpretation\ of\ quantum\ entanglement
characterized\ by\ the\ purity%
\begin{equation}
\mathcal{P}=%
{\displaystyle\iiiint}
\psi\left(  k_{1},k_{2},t\right)  \psi\left(  k_{3},k_{4},t\right)  \psi
^{\ast}\left(  k_{1},k_{4},t\right)  \psi^{\ast}\left(  k_{3},k_{2},t\right)
dk_{1}dk_{2}dk_{3}dk_{4}, \label{purity2}%
\end{equation}
we employ our \emph{second conjecture} whereby we use IG to model the
scattering interaction by patching together two charts, each belonging to a
different Gaussian statistical manifolds, one without correlation (pre
collision) and the other with correlation (post collision). The two models can
be represented by means of $p_{\text{IG}}^{\text{(uncorr)}}\left(  x,y|\mu
_{x},\mu_{y},\sigma\right)  $ and $p_{\text{IG}}^{\text{(corr)}}\left(
x,y|\mu_{x},\mu_{y},\sigma;\rho\right)  $ with associated statistical
manifolds $\mathcal{M}^{\text{(uncorr)}}$ and $\mathcal{M}^{\text{(corr)}}$,
respectively. The two charts belonging to the correlated and uncorrelated
Gaussian statistical manifolds are patched together by joining the sets of
geodesic curves associated with each manifold at the junction\textbf{\ }%
$\tau=0$. In particular, the set of geodesic curves defined when $\tau
<0$\ (pre collision) for the\ uncorrelated model is joined to the set defined
when $\tau\geq0$\ (post collision) for the correlated model. As a consequence
of our second conjecture, we are able to uncover an interesting quantitative
connection between the correlation coefficient $\rho$ and the scattering
potential $V(x)$ on the one hand,\ and purity $\mathcal{P}$ and the IGC
$\mathcal{C}$ on the other. Full details of the implementation of our second
conjecture can be found in \cite{kim1, kim2}.

By direct computation of the integral in Eq\textbf{.} (\ref{purity2}), we
obtain%
\begin{equation}
\mathcal{P}\approx1-\frac{2\left(  2k_{\mathrm{o}}^{2}+\sigma_{k\mathrm{o}%
}^{2}\right)  R_{\mathrm{o}}\sqrt{\Sigma}}{\sqrt{\pi}},
\end{equation}
where $\Sigma=4\pi\left\vert f\left(  k_{\mathrm{o}}\right)  \right\vert ^{2}$
is the scattering cross-section. Next we seek to determine $\Sigma$ by
obtaining an expression for $f\left(  k_{\mathrm{o}}\right)  $. Under the
assumption that the two particles are well separated both initially (before
collision) and finally (after collision), and further assuming that the
colliding Gaussian wave packets are very narrow in momentum space
($\sigma_{k_{\mathrm{o}}}\ll1$ such that the phase shift can be treated as a
constant $\theta\left(  k_{\mathrm{o}}\right)  $), we deduce that the
scattering potential is given by%
\begin{equation}
V=\rho\mathcal{E}=\rho\frac{\hbar^{2}k_{\mathrm{o}}^{2}}{2\mu}=\rho
\frac{p_{\mathrm{o}}^{2}}{2\mu}.
\end{equation}
With the potential determined, we obtain the scattering phase shift
\begin{equation}
\tan\theta_{\mathrm{o}}\overset{k_{\mathrm{o}}L=p_{\mathrm{o}}L/\hbar\ll
1}{\approx}\theta_{\mathrm{o}}\approx-\frac{\rho\left(  k_{\mathrm{o}%
}L\right)  ^{3}}{3}=-\frac{2\mu Vk_{\mathrm{o}}L^{3}}{3\hbar^{2}}=-\frac{2\mu
Vp_{\mathrm{o}}L^{3}}{3\hbar^{3}}, \label{eq:phase_shift_V}%
\end{equation}
which is in perfect agreement with \cite{Mishima}\textbf{\ }(and not found in
\cite{Wang}) where standard Schr\"{o}dinger's quantum dynamics was employed.
This result is significant because it allows to state that our conjecture is
also physically motivated. As the scattering potential has been determined, so
too can the scattering amplitude be obtained. To this end, we write
\begin{equation}
f\left(  k_{\mathrm{o}}\right)  =\frac{e^{i\theta_{\mathrm{o}}}\sin
\theta_{\mathrm{o}}}{k_{\mathrm{o}}}\approx\frac{\theta_{\mathrm{o}}%
}{k_{\mathrm{o}}}\approx-a_{\mathrm{s}} \label{eq:f01}%
\end{equation}
for low energy $s$-wave scattering, $k_{\mathrm{o}}L=p_{\mathrm{o}}L/\hbar
\ll1$. Then the squared modulus of Eq\textbf{.} (\ref{eq:f01}), by means of
Eq. (\ref{eq:phase_shift_V}), reads
\begin{equation}
\left\vert f\left(  k_{\mathrm{o}}\right)  \right\vert ^{2}\approx\frac
{\theta_{\mathrm{o}}^{2}}{k_{\mathrm{o}}^{2}}\approx\frac{\rho^{2}%
k_{\mathrm{o}}^{4}L^{6}}{9}=\frac{4\mu^{2}V^{2}L^{6}}{9\hbar^{4}}\approx
a_{\mathrm{s}}^{2}. \label{eq:f02}%
\end{equation}
Thus, we finally obtain the scattering cross section:
\begin{equation}
\Sigma=4\pi\left\vert f\left(  k_{\mathrm{o}}\right)  \right\vert ^{2}%
\approx\frac{4\pi\rho^{2}k_{\mathrm{o}}^{4}L^{6}}{9}=\frac{16\pi\mu^{2}%
V^{2}L^{6}}{9\hbar^{4}}\approx4\pi a_{\mathrm{s}}^{2}.
\label{eq:cross-section}%
\end{equation}
Having found the scattering cross-section, we can recast the purity of the
post-collisional two-particle wave function and the correlation coefficient
as
\begin{equation}
\mathcal{P}\approx1-\frac{4\rho k_{\mathrm{o}}^{2}\left(  2k_{\mathrm{o}}%
^{2}+\sigma_{k_{\mathrm{o}}}^{2}\right)  R_{\mathrm{o}}L^{3}}{3}=1-\frac{8\mu
V\left(  2k_{\mathrm{o}}^{2}+\sigma_{k_{\mathrm{o}}}^{2}\right)
R_{\mathrm{o}}L^{3}}{3\hbar^{2}}, \label{eq:P1}%
\end{equation}
and
\begin{equation}
\rho=\frac{V}{\mathcal{E}}=\frac{2\mu V}{\hbar^{2}k_{\mathrm{o}}^{2}}%
\approx\frac{3\sqrt{\Sigma}}{2\sqrt{\pi}k_{\mathrm{o}}^{2}L^{3}}\approx
\frac{3a_{\mathrm{s}}}{k_{\mathrm{o}}^{2}L^{3}}\text{,} \label{corr2}%
\end{equation}
respectively. By use of our first conjecture $\rho=\rho_{\mathrm{QM}}$, with
$\rho_{\mathrm{QM}}\overset{\text{def}}{=}\sqrt{8\left(  2k_{\mathrm{o}}%
^{2}+\sigma_{k\mathrm{o}}^{2}\right)  R_{\mathrm{o}}a_{\mathrm{s}}}$, together
with Eq. (\ref{eq:cross-section})\ and Eq.\textbf{ }(\ref{corr2}), we are able
to determine the scattering \textit{potential density}
\begin{equation}
\frac{V}{L^{3}}=\frac{4\hbar^{2}k_{\mathrm{o}}^{4}\left(  2k_{\mathrm{o}}%
^{2}+\sigma_{k\mathrm{o}}^{2}\right)  R_{\mathrm{o}}}{3\mu}, \label{Vdensity}%
\end{equation}
as well as the scattering length,
\begin{equation}
a_{\mathrm{s}}\approx2\mu VL^{3}/3\hbar^{2}.
\end{equation}
This result for the scattering length agrees with equation $(41)$\ of
\cite{Mishima}. Equation (\ref{eq:P1}) demonstrates that the purity
$\mathcal{P}$ can be expressed in terms of physical quantities such as the
scattering potential height $V$ and range $L$ together with the initial
quantities $k_{\mathrm{o}}$, $\sigma_{\mathrm{o}}$ and $R_{\mathrm{o}}$.
This\textbf{ }result constitutes the second significant finding obtained
within our hybrid approach which explains how the entanglement strength is
controlled by the interaction potential height $V$ and the incident kinetic
energy $\mathcal{E}$ of the two-particle system. The role played by $\rho$ in
the quantities $\mathcal{P}$ and $V$ seem to\textbf{ }suggest that the
physical information about quantum entanglement is encoded in the covariance
term $\mathrm{Cov}\left(  k_{1},k_{2}\right)  \overset{\text{def}}%
{=}\left\langle k_{1}k_{2}\right\rangle -\left\langle k_{1}\right\rangle
\left\langle k_{2}\right\rangle $ appearing in the definition of the
correlation coefficient $\rho$.

\subsubsection{Information Geometric Complexity of Entangled Gaussian
Wave-Packets}

The line element of the Fisher-Rao information metric\textbf{ }on the
statistical manifold $\mathcal{M}^{\text{(corr)}}$ induced by $p_{\text{IG}%
}^{\text{(corr)}}\left(  x,y|\mu_{x},\mu_{y},\sigma;r\right)  $ is given by%
\begin{equation}
ds_{\mathcal{M}^{\text{(corr)}}}^{2}=\frac{1}{\sigma^{2}}\left(  \frac
{1}{1-\rho^{2}}d\mu_{x}^{2}+\frac{1}{1-\rho^{2}}d\mu_{y}^{2}-\frac{2\rho
}{1-\rho^{2}}d\mu_{x}d\mu_{y}+4d\sigma^{2}\right)  .
\end{equation}
Note that $\mathcal{M}^{\text{(uncorr)}}=\left\{  \left.  p_{\text{IG}%
}^{\text{(uncorr)}}\left(  x,y|\mu_{x},\mu_{y},\sigma\right)  \right\vert
p_{\text{IG}}^{\text{(uncorr)}}\left(  x,y|\mu_{x},\mu_{y},\sigma\right)
\geq0\right\}  $, and $ds_{\mathcal{M}^{\text{(uncorr)}}}^{2}=ds_{\mathcal{M}%
^{\text{(corr)}}}^{2}\left(  \rho\rightarrow0\right)  $. The IGC is determined
to be%
\begin{equation}
\mathcal{C}_{\mathcal{M}^{\text{(corr)}}}\left(  \tau;\rho\right)  =\frac
{8}{\lambda_{\mathcal{M}}}\sqrt{\frac{1-\rho}{1+\rho}}\left[  -\frac{3}%
{4}\lambda_{\mathcal{M}}+\frac{1}{4}\frac{\sinh\left(  \lambda_{\mathcal{M}%
}\tau\right)  }{\tau}+\frac{\tanh\left(  \frac{1}{2}\lambda_{\mathcal{M}}%
\tau\right)  }{\tau}\right]  . \label{complexity}%
\end{equation}
Similarly, $\mathcal{C}^{\text{(uncorr)}}=\mathcal{C}^{\text{(corr)}}\left(
\rho\rightarrow0\right)  $ represents the IGC on $\mathcal{M}^{\text{(uncorr)}%
}$.

The technical details that will be omitted in what follows may be found in
\cite{kim1, kim2}. By direct computation, the post-collision\textbf{ }IGE is
found to be%
\begin{equation}
\mathcal{S}_{\mathcal{M}^{\text{(corr)}}}\left(  \tau;\rho\right)
=\lambda_{\mathcal{M}}\tau-\log\left(  \lambda_{\mathcal{M}}\tau\right)
+\frac{1}{2}\log\left(  \frac{1-\rho}{1+\rho}\right)  . \label{eq:entrp_r}%
\end{equation}
For uncorrelated Gaussian statistical models, the IGE is given by
$\mathcal{S}_{\mathcal{M}^{\text{(uncorr)}}}\left(  \tau;0\right)
=\mathcal{S}_{\mathcal{M}^{\text{(corr)}}}\left(  \tau;\rho\rightarrow
0\right)  $. For the specific case being considered in this work, the IG
analogue of the KS-entropy is determined to be
\begin{equation}
h_{\mathcal{M}}^{\text{KS}}\approx2A_{\mathrm{o}}=\lambda_{\mathcal{M}}.
\label{hks1}%
\end{equation}
The KS-entropy is related to the coarse-grained Boltzmann entropy\textbf{\ }%
according to \cite{Ropotenko},%
\begin{equation}
\mathcal{S}_{\text{B}}\left(  t\right)  =h^{\text{KS}}t. \label{SBol}%
\end{equation}
By means of Eq. (\ref{hks1}) we observe that the IGE is related to the
KS-entropy in a similar manner as the coarse-grained Boltzmann entropy in
Eq\textbf{. }(\ref{SBol}), as seen in the following:
\begin{equation}
\mathcal{S}_{\mathcal{M}^{\text{(uncorr)}}}\left(  \tau;0\right)  \approx
h_{\mathcal{M}}^{\text{KS}}\tau. \label{new1}%
\end{equation}
By comparing the asymptotic expressions of the IGCs in the presence and
absence of correlations, respectively, we obtain the IGC ratio%
\begin{equation}
\frac{\mathcal{C}_{\mathcal{M}^{\text{(corr)}}}\left[  \mathcal{D}_{\theta
}^{\text{(geodesic)}}\left(  \tau;\rho\right)  \right]  }{\mathcal{C}%
_{\mathcal{M}^{\text{(uncorr)}}}\left[  \mathcal{D}_{\theta}%
^{\text{(geodesic)}}\left(  \tau;0\right)  \right]  }=\sqrt{\frac{1-\rho
}{1+\rho}}. \label{eq:v-rel}%
\end{equation}
From Eq\textbf{. }(\ref{eq:entrp_r}) we are also able to determine,
\begin{equation}
\mathcal{S}_{\mathcal{M}^{\text{(corr)}}}\left(  \tau;\rho\right)
-\mathcal{S}_{\mathcal{M}^{\text{(uncorr)}}}\left(  \tau;0\right)  =\frac
{1}{2}\log\left(  \frac{1-\rho}{1+\rho}\right)  . \label{s-rel}%
\end{equation}
From Eqs. (\ref{eq:v-rel}) and (\ref{s-rel}) we find that both the IGC and IGE
decrease in presence of correlations.\ Specifically, the former decreases by
the factor $\sqrt{\frac{1-\rho}{1+\rho}}<1$ for $\rho>0$ whereas\ the latter
decreases by\textbf{ }$\frac{1}{2}\log\left(  \frac{1-\rho}{1+\rho}\right)
<0$ for $\rho>0$. Furthermore, inspection of Eq. (\ref{eq:v-rel}) confirms
that an increase in the correlational structure among the macrovariables of a
system implies a reduction in the complexity of the corresponding geodesic
information flows on the underlying statistical manifold \cite{CM,
cafaro-mancini, carlo-PD} of said system. Stated otherwise, drawing
macroscopic predictions is easier in the presence of correlations than in
their absence.

From Eq. (\ref{complexity}) we deduce%
\begin{equation}
\rho=\frac{V}{\mathcal{E}}=\frac{\Delta\mathcal{C}^{2}}{\mathcal{C}%
_{\text{total}}^{2}}\text{,} \label{due2}%
\end{equation}
with
\begin{equation}
\Delta\mathcal{C}^{2}\overset{\text{def}}{=}\left[  \mathcal{C}%
^{\text{(uncorr)}}\right]  ^{2}-\left[  \mathcal{C}^{\text{(corr)}}\right]
^{2}\text{, and }\mathcal{C}_{\text{total}}^{2}\overset{\text{def}}{=}\left[
\mathcal{C}^{\text{(uncorr)}}\right]  ^{2}+\left[  \mathcal{C}^{\text{(corr)}%
}\right]  ^{2}\text{.}%
\end{equation}
By combining Eqs\textbf{. }(\ref{eq:P1}) and (\ref{complexity}) it follows
that%
\begin{equation}
\mathcal{P}\approx1-\eta_{\mathcal{C}}\cdot\frac{\left(  \Delta\mathcal{C}%
\right)  ^{2}}{\mathcal{C}_{\text{total}}^{2}}\text{ where }\eta_{\mathcal{C}%
}\overset{\text{def}}{=}\frac{8}{3}k_{\mathrm{o}}^{2}\left(  2k_{\mathrm{o}%
}^{2}+\sigma_{k_{\mathrm{o}}}^{2}\right)  R_{\mathrm{o}}L^{3}. \label{ultima}%
\end{equation}
A new quantitative relation between quantum entanglement and IGC was uncovered
in Eq. (\ref{ultima}). From this relation it is evident that our system
becomes perfectly pure (i.e. $\mathcal{P}=1$) for vanishing complexity. On the
other hand, as the complexity increases from zero, the mixedness of the system
increases, causing the purity to decrease from unity. The appearance
of\ correlation terms leads to the compression of the correlated IGC by the
fraction $\sqrt{\frac{1-\rho}{1+\rho}}$. From this result we conclude that the
IGC decreases when the quantum wave-packets comprising our system becomes entangled.

By means of the hybrid approach employed in this application, we obtained
results which coincide with those in \cite{Wang} (mathematical support to our
conjecture)\ in addition to those in \cite{Mishima}\ (physical support to our
conjecture). While these results are not new, they nonetheless motivate the
utility of our theoretical modeling scheme in the study of entanglement.

\section{Concluding Remarks}

In this article, we presented a theoretical modeling scheme that combines IG
techniques with inductive inference methods. This modeling scheme allows to
describe the macroscopic behavior of complex systems in terms of the
microscopic degrees of freedom of the system. After the MrE and IG formalisms
were reviewed, particular emphasis was placed on our information geometric
measures of complexity, namely the IGC and the IGE. Ten illustrative examples
of this modeling scheme were presented.

\begin{itemize}
\item Application 1: For our first example we investigated a Gaussian
statistical model describing an arbitrary system with $l$ uncorrelated degrees
of freedom. In this case it was determined that the IGE of this statistical
model exhibits linear growth characteristics. This asymptotic linear growth of
the IGE may be considered the IG analogue of the von Neumann entropy growth
introduced by Zurek and Paz.
\end{itemize}

Our second and third applications focused on bivariate and trivariate Gaussian
statistical models, respectively, each admitting correlations among the
microstates of each system.

\begin{itemize}
\item Application 2: In the bivariate case, the ratio of the IGCs in the
presence and absence of correlations was found to be a monotonic increasing
function of the correlation parameter $\rho$ over the range of values $\left(
-1,1\right)  $. From this result we conclude that entropic inferences on two
Gaussian distributed microstates is implemented in a less (more) efficient
manner when the two microstates are negatively (positively) correlated than in
the absence of correlations.

\item Application 3\textbf{:} The trivariate scenario admits three cases, with
each case corresponding to the three possible choices of the covariance
matrix. We termed these three cases \emph{weak}, \emph{mildly weak} and
\emph{strong}, with each designation corresponding to a minimal, intermediate
(i.e. neither minimal nor maximal) and maximal correlations among the
microstates of the system respectively. In the weak case, the IGC ratio
exhibits a monotonic behavior in the correlation parameter $\rho\in$ $\left(
-1,1\right)  $. For the mildly weak case, the IGC ratio exhibits a
non-monotonic behavior in $\rho\in\left(  -\frac{\sqrt{2}}{2},\frac{\sqrt{2}%
}{2}\right)  $, assuming a maxima at $\rho=\frac{1}{2}$ and vanishing at
$\rho=\frac{\sqrt{2}}{2}$. For the strong case, the IGC ratio exhibits a
monotonic behavior in $\rho\in$ $\left(  -\frac{1}{2},1\right)  $. In the
latter case, contrary to the mildly weak case, the IGC ratio cannot be zero at
the extrema of the range. This behavior is similar to the geometric
frustration phenomena that occurs in the presence of loops. Finally,
comparison of the IGC ratios for the maximally correlated trivariate and
bivariate cases leads to conclude that carrying out entropic inferences on a
higher-dimensional manifold in the presence of anti-correlations is less
complex than on a lower-dimensional manifold. The converse is true in the
presence of positive-correlations.

\item Application 4: In our forth example, we sought the understanding of the
asymptotic temporal behavior of the IGC of a three-dimensional microcorrelated
Gaussian statistical model. In this scenario, it was observed that the
presence of microcorrelations result in an asymptotic power law decay of the
IGC. This decay leads in effect, to the emergence of an asymptotic information
geometric compression of the statistical macrostates explored by the system
during its evolution, at a faster rate compared to the case of vanishing correlations.

\item Application 5: Our fifth example examined the effect upon a statistical
model due to the introduction of embedding constraints on the macrostates of
the system. In presence of such constraints, it is observed that the
introduction of embedding constraints lead to the emergence of an asymptotic
compression of the statistical macrostates explored by the system as it
evolves, with this compression occurring at a faster rate than that observed
in absence of embedding constraints. Although arising through radically
different mechanisms, the results of Application 4 and 5 both provide
quantitative evidence that the information geometric complexity of a
statistical systems decreases in presence of correlational structures of
either macroscopic or microscopic nature.

The sixth and seventh examples focused respectively, on the dimensional
reduction arising from quantum-like constraints being imposed on the
macrostates of both $3$-dimensional uncorrelated ($3Du$) and $3$-dimensional
correlated ($3Dc$) Gaussian statistical systems.

\item Application 6: In this application, it is found that when a quantum-like
constraint reminiscent of Heisenberg's minimum uncertainty relation is imposed
on the macrostates of a $3Du$ Gaussian model, the IGE of the corresponding
$2Du$ system is attenuated relative to the $3Du$ case.

\item Application 7: In this example, it is found that the introduction of
quantum-like constraints on the macrostates of a $3Dc$ Gaussian model results
in a corresponding $2Dc$ model that exhibits no asymptotic change in the IGE
relative to the $2Du$ case. This comparison between the uncorrelated and
correlated $2D$ Gaussian models demonstrate that further constraining the
$2Du$ with a quantum-like constraint does not result in further global
softening of complexity. By contrast, the IGC of the correlated $2Dc$ model
diverges as the correlation coefficient approaches unity.

\item Application 8: In our eighth example, we studied the IG model of an
ensemble of random frequency, macroscopic, inverted harmonic oscillators. Our
analysis led to the result that the IGE of the model grew linearly in affine
time. This may be viewed as an IG analogue of the Zurek-Paz quantum chaos
criterion of asymptotic linear entropy growth, and effectively extends the
result of Zurek-Paz to an arbitrary set of anisotropic inverted harmonic
oscillators in the classical information-geometric setting.

\item Application 9: For our ninth application, we investigated the
statistical manifolds induced by classical probability distributions commonly
used in the study of regular and chaotic quantum energy level statistics. As a
result of our analysis, it was determined that the IGE associated with regular
and chaotic spin chains displayed asymptotic logarithmic and linear growth
characteristics in affine time, respectively. These results may be seen as IG
analogues of the regular and chaotic entanglement entropies arising in quantum
energy level statistics, since the asymptotic behavior of these latter
entropies are known to grow logarithmically and linearly in time, respectively.

\item Application 10\textbf{:} Our tenth and final example made use of our
modeling scheme to describe the scattering-induced quantum entanglement
between two Gaussian minimum uncertainty wave-packets. It was found that the
correlation coefficient $\rho$ can be viewed as the ratio of the potential to
kinetic energy of the system. This result constitutes an explicit connection
between correlations and physical observables (the macrostate $p_{\mathrm{o}}$
in this case). When $\rho\neq0$ the wave-packets experience the effect of a
repulsive potential; the magnitude of the wave vectors (momenta) decreases
relative to their corresponding uncorrelated value. Relative to the
uncorrelated case, it was determined that for scenario in which $\rho>0$, the
IGC is compressed by the factor $\sqrt{\frac{1-\rho}{1+\rho}}<1$, whereas\ the
IGE decreases by the amount $\frac{1}{2}\log\left(  \frac{1-\rho}{1+\rho
}\right)  <0$. Finally, it was observed that the uncorrelated IGE is related
to the KS-entropy in the same functional manner as the coarse-grained
Boltzmann entropy is related to the KS-entropy.
\end{itemize}

At this juncture, it is worth noting that although the first three
applications introduced in this article analyze Gaussian models of arbitrary
nature, we emphasize that an understanding of such systems is nevertheless
quite useful since they facilitate the construction of more physically
motivated scenarios such as our IG models of anisotropic, inverted harmonic
oscillators and scattering induced quantum entanglement presented in examples
eight and ten, respectively. These two examples lead to a connection between
our IG modeling scheme and actual physical systems by means of the former
reproducing the known results of linear von Neumann entropy growth attributed
to Zurek-Paz \cite{zurek1} in the case of example eight, and to the results
found in Mishima \cite{Mishima} for the scattering phase shift and scattering
length in the case of example ten. In example nine, we utilized our
theoretical modeling scheme to describe integrable (by means of a composite
Poisson and Exponential model) and chaotic (by means of a composite
Wigner-Dyson and Gaussian model) energy level statistics associated with a
quantum antiferromagnetic Ising spin chain leading to an IG result that is
coincident with the known entanglement entropy of each scenario. Specifically,
the IGE exhibits asymptotic logarithmic growth in the former case, while the
IGE exhibits asymptotic linear growth in the latter case. These results are
consistent with the known results that integrable and chaotic quantum
antiferromagnetic Ising chains are characterized by asymptotic logarithmic and
linear growths of their entanglement entropies \cite{Prosen}, respectively.
Finally, building upon example four and five, we find that in example six, Eq.
(\ref{IGEfinal}) quantitatively shows that the IGE\ is attenuated when
approaching the\textbf{ }$2Du$\textbf{ }case from the\textbf{ }$3Du$\textbf{
}case via the introduction of the macroscopic constraint in Eq. (\ref{MC})
that is reminiscent of Heisenberg's minimum uncertainty relation. In the same
vein of our work in \cite{OSID}, a recent investigation concludes that quantum
mechanics can reduce the statistical complexity of classical models \cite{nc}.
It is our hope that, based on the above discussion, the connection between our
theoretical modeling scheme and physics is made more transparent.

In conclusion, we would like to outline four possible lines of research for
future investigations:

\begin{itemize}
\item Describe the role of thermodynamics within the IGAC. For instance,
thermodynamics is known to play a prominent role in the entropic analysis of
chaotic dynamics \cite{beck}.

\item Better understand the role of thermodynamics as a possible bridge among
different complexity measures.

\item Further develop our understanding of how the IGE\ relates to the
Kolmogorov-Sinai dynamical entropy, the coarse-grained Boltzmann entropy and
the von Neumann entropy depending on the nature of the system being modeled;
and finally,

\item Explore in a more substantial manner the relationship between our IGE
and the intrinsic complexity appearing in the work of Rodriguez
\cite{rodriguez04} as well as in the works by Myung, Balasubramanian and Pitt
\cite{minghia1, vijay, minghia2}.
\end{itemize}

We are aware of several issues that remain unsolved within the IGAC and
further development of the framework remains to be done. Nevertheless, we are
gratified\ that our theoretical modeling scheme is gaining attention within
the community. Indeed, there appears to be an increasing number of scientists
who are actively engaged in research that either makes use of, or is related
to, the theoretical framework described in the present article \cite{peng,
peng2, r1,r2,r3,FMP,r5,r6,r7,r8,r9,r10,r11,r12,r13,r14,r15,r16,r17,
r18,r18a,r19}.

\begin{acknowledgments}
The authors are grateful to Domenico Felice for helpful discussions in the
early version of this manuscript. Finally, C. C. acknowledges the hospitality
of the United States Air Force Research Laboratory in Rome (New York) where
part of his contribution to this work was completed.
\end{acknowledgments}


\begin{thebibliography}{999}                                                                                              %


\bibitem {jimmy}J. P. Crutchfield and B. S. McNamara, \emph{Equations of
motions from a data series}, Complex Systems \textbf{1}, 417 (1987).

\bibitem {minghia1}I. J. Myung, V. Balasubramanian, and M. A. Pitt,
\emph{Counting probability distributions: differential geometry and model
selection}, Proc. Natl Acad. Sci. \textbf{97}, 11170 (2000).

\bibitem {vijay}V. Balasubramanian, \emph{A geometric formulation of Occam's
razor for inference of parametric distributions}, arXiv:adap-org/9601001 (1996).

\bibitem {minghia2}V. Balasubramanian, \emph{Statistical inference, Occam's
razor and statistical mechanics on the space of probability distributions},
Neural Computation \textbf{9}, 268 (1997).

\bibitem {toby}T. S. Cubitt, J. Eisert, and M. W. Wolf, \emph{Extracting
dynamical equations from experimental data is NP hard}, Phys. Rev. Lett.
\textbf{108}, 120503 (2012).

\bibitem {gell-mann}M. Gell-Mann, \emph{What is complexity?}, Complexity
\textbf{1}, 1 (1995).

\bibitem {james1}D. P. Feldman and J. P. Crutchfield, \emph{Measures of
complexity: why?}, Phys. Lett. \textbf{A238}, 244 (1998).

\bibitem {james2}J. P. Crutchfield and K. Young, \emph{Inferring statistical
complexity}, Phys. Rev. Lett. \textbf{63}, 105 (1989).

\bibitem {landauer}R. Landauer, \emph{A simple measure of complexity}, Nature
\textbf{336}, 306 (1988).

\bibitem {romano}R. Romano and P. van Loock, \emph{Quantum control of noisy
channels}, arXiv:quant-ph/0811.3014 (2008).

\bibitem {lloyd1}S. Lloyd and H. Pagels, \emph{Complexity as thermodynamic
depth}, Ann. Phys. \textbf{188}, 186 (1988).

\bibitem {bennett-f}C. H. Bennett, \emph{On the nature and origin of
complexity in discrete, homogeneous, locally-interacting systems}, Foundations
of Physics \textbf{16}, 585 (1986).

\bibitem {bennett-book}C. H. Bennett, \emph{How to define complexity in
physics and why}, in Complexity, Entropy, and the Physics of Information:
Proceedings of the Santa Fe Institute Workshop, ed. by W. H. Zurek (1989).

\bibitem {lloyd2013}S. Lloyd, \emph{Core-halo instability in dynamical
systems}, arXiv:nlin.CD/1302.3199 (2013).

\bibitem {jie}J. Sun, E. M. Bollt and T. Nishikawa, \emph{Judging model
reduction of complex systems}, Phys. Rev. \textbf{E83}, 046125 (2011).

\bibitem {jaynes1}E. T. Jaynes, \emph{Macroscopic prediction}, in Complex
Systems and Operational Approaches in Neurobiology, Physics, and Computers,
ed. by H. Haken, Springer, Berlin (1985).

\bibitem {caticha-ED}A. Caticha, \emph{Entropic dynamics}, AIP Conf. Proc.
\textbf{617}, 302 (2002).

\bibitem {caticha-giffin}A. Caticha and A. Giffin, \emph{Updating
probabilities}, AIP Conf. Proc. \textbf{872}, 31 (2006).

\bibitem {amari-japan}S. Amari and H. Nagaoka, $\emph{Methods}$ $\emph{of}$
$\emph{information}$ $\emph{geometry}$, Oxford University Press (2000).

\bibitem {PR}L. Casetti, M. Pettini and E. G. D. Cohen, \emph{Geometric
approach to Hamiltonian dynamics and statistical mechanics}, Phys. Rep.
\textbf{337}, 237 (2000).

\bibitem {chicone}C. Chicone and B. Mashhoon, \emph{The generalized Jacobi
equation}, Class. Quantum Grav. 19 4231-4248 (2002).

\bibitem {caticha-cafaro}A. Caticha and C. Cafaro, \emph{From information
geometry to Newtonian dynamics}, AIP Conf. Proc. \textbf{954}, 165 (2007).

\bibitem {caticha-jpa}A. Caticha, \emph{Entropic dynamics, time and quantum
theory}, J. Phys. \textbf{A}: Math. Theor. \textbf{44}, 225303 (2011).

\bibitem {giffin-caticha}A. Giffin and A. Caticha, \emph{Updating
probabilities with data and moments}, AIP Conf. Proc. \textbf{954}, 74 (2007).

\bibitem {giffin16}A. Giffin, C. Cafaro, and S. A. Ali, \emph{Application of
the maximum relative entropy method to the physics of ferromagnetic
materials}, Physica \textbf{A455}, 11 (2016).

\bibitem {carlo-tesi}C. Cafaro, \emph{The information geometry of chaos}, PhD
Thesis, State University of New York at Albany, Albany-NY, USA (2008).

\bibitem {casetti}L. Casetti, C. Clementi, and M. Pettini, \emph{Riemannian
theory of Hamiltonian chaos and Lyapunov exponents}, Phys. Rev. \textbf{E54},
5969 (1996).

\bibitem {di bari}M. Di Bari and P. Cipriani, \emph{Geometry and chaos on
Riemann and Finsler manifolds}, Planet. Space Sci. \textbf{46}, 1543 (1998).

\bibitem {jacobi}C. G. J. Jacobi, \emph{Vorlesungen uber dynamik}, Reimer,
Berlin (1866).

\bibitem {fisher}R.A. Fisher, \emph{Theory of statistical estimation}, Proc.
Cambridge Philos. Soc. \textbf{122}, 700 (1925).

\bibitem {rao}C.R. Rao, \emph{Information and accuracy attainable in the
estimation of statistical parameters}, Bull. Calcutta Math. Soc. \textbf{37},
81 (1945).

\bibitem {ARIEL}A. Caticha, \emph{Lectures on probability, entropy and
statistical physics}, in the 28-th International Workshop on Bayesian
Inference and Maximum Entropy Methods in Science and Engineering, Brazil (2008).

\bibitem {defelice}F. De Felice and J. S. Clarke, \emph{Relativity on curved
manifolds}, Cambridge University Press (1990).

\bibitem {carmo}M. P. do Carmo, \emph{Riemannian geometry}, Birkhauser, Boston (1992).

\bibitem {carlopre16}C. Cafaro and S. A. Ali, \emph{Maximum caliber inference
and the stochastic Ising model}\textbf{, }Phys. Rev. \textbf{E94}, 052145 (2016).

\bibitem {tutorial}A. Caticha, \emph{Entropic inference and the foundations of
physics}, USP Press, Sao Paulo, Brazil (2012).

\bibitem {carlo-PD}C. Cafaro and S. A. Ali, \emph{Jacobi fields on statistical
manifolds of negative curvature}, Physica \textbf{D234}, 70 (2007).

\bibitem {carlo-AMC}C. Cafaro, A. Giffin, S. A. Ali and D.-H. Kim,
\emph{Reexamination of an information geometric construction of entropic
indicators of complexity}, Appl. Math. Comput. \textbf{217}, 2944 (2010).

\bibitem {rodriguez04}C. C. Rodriguez, \emph{The volume of bitnets}, AIP Conf.
Proc. \textbf{735}, 555 (2004).

\bibitem {caves}C. M. Caves and R. Schack, \emph{Unpredictability,
information, and chaos}, Complexity \textbf{3}, 46-57 (1997); A. J. Scott, T.
A. Brun, C. M. Caves, and R. Schack, \emph{Hypersensitivity and chaos
signatures in the quantum baker's map}, J. Phys. \textbf{A39}, 13405 (2006).

\bibitem {alekseev}V. M. Alekseev and M. V. Yakobson, \emph{Symbolic dynamics
and hyperbolic dynamic systems}, Phys. Reports \textbf{75}, 287 (1981).

\bibitem {kolmogorov-c}A. N. Kolmogorov, \emph{Three approaches to the
quantitative definition of information}, Probl. Inf. Transm. (USSR)
\textbf{1}, 4 (1965); A. N. Kolmogorov, \emph{Logical basis for information
theory and probability theory}, IEEE Trans. Inf. Theory,\textbf{\ IT14}, 662
(1968); T. M. Cover and J. A. Thomas, \emph{Elements of information theory},
John Wiley and Sons, Inc. (2006).

\bibitem {pesin}Y. Pesin, \emph{Characteristic Lyapunov exponents and smooth
ergodic theory}, Russian Mathematics Survey \textbf{32}, 55 (1977).

\bibitem {benattaman}F. Benatti, \emph{Classical and quantum entropies:
dynamics and Information}, in \textit{Entropy}, Princeton University Press
(2003); T. M. Cover and J. A. Thomas, \emph{Elements of information theory},
John Wiley and Sons, Inc. (2006).

\bibitem {PR9}C. Cafaro, S. A. Ali, and A. Giffin, \emph{An application of
reversible entropic dynamics on curved statistical manifolds}, in Bayesian
Inference and Maximum Entropy Methods in Science and Engineering, AIP Conf.
Proc. \textbf{872}, 243 (2006).

\bibitem {PR7}C. Cafaro, \emph{Information geometry and chaos on negatively
curved statistical manifolds}, in Bayesian Inference and Maximum Entropy
Methods in Science and Engineering, AIP Conf. Proc. \textbf{954}, 175 (2007).

\bibitem {PR6}C. Cafaro, \emph{Recent theoretical progress on an information
geometrodynamical approach to chaos}, in Bayesian Inference and Maximum
Entropy Methods in Science and Engineering, AIP Conf. Proc. \textbf{1073}, 16 (2008).

\bibitem {IG4}S. A. Ali, C. Cafaro, A. Giffin, and D.-H. Kim, \emph{Complexity
characterization in a probabilistic approach to dynamical systems through
information geometry and inductive inference}, Physica Scripta \textbf{85},
025009 (2012).

\bibitem {PR1}C. Cafaro\textbf{, }\emph{Information geometric complexity of
entropic motion on curved statistical manifolds}, Proceedings of the 12th
Joint European Thermodynamics Conference, JETC 2013, Eds. M. Pilotelli and
G.P. Beretta (ISBN 978-88-89252-22-2, Snoopy, Brescia, Italy, 2013), pp. 110-118.

\bibitem {carlo-IJTP}C. Cafaro, \emph{Information-geometric indicators of
chaos in Gaussian models on statistical manifolds of negative Ricci
curvature}, Int. J. Theor. Phys. \textbf{47}, 2924 (2008).

\bibitem {Zurek}W. H. Zurek, S. Habib and J. P. Paz, \emph{Coherent states via
decoherence}, Phys. Rev. Lett. \textbf{70}, 1187 (1993).

\bibitem {roz}Y. A. Rozanov, \emph{Probability theory: a concise course},
Dover Publications, New York (1977).

\bibitem {CNM-entropy}D. Felice, C. Cafaro, and S. Mancini,\emph{\ Information
geometric complexity of a trivariate Gaussian statistical model}, Entropy
\textbf{16}, 2944 (2014).

\bibitem {SM99}J. F. Sadoc and R. Mosseri, \emph{Geometrical frustration},
Cambridge University Press (2006).

\bibitem {FMP}D. Felice, S. Mancini and M. Pettini, \emph{Quantifying networks
complexity from information geometry viewpoint}, J. Math. Phys. \textbf{55},
043505 (2014).

\bibitem {carloPA2010}S. A. Ali, C. Cafaro, D.-H. Kim and S. Mancini,
\emph{The effect of microscopic correlations on the information geometric
complexity of Gaussian statistical models}, Physica \textbf{A389}, 3117 (2010).

\bibitem {CM}C. Cafaro and S. Mancini, \emph{On the complexity of statistical
models admitting correlations}, Phys. Scr. \textbf{82}, 035007 (2010).

\bibitem {cafaro-mancini}C. Cafaro and S. Mancini, \emph{Quantifying the
complexity of geodesic paths on curved statistical manifolds through
information geometric entropies and Jacobi fields}, Physica \textbf{D240}, 607 (2011).

\bibitem {PR3}C. Cafaro, A. Giffin, C. Lupo, and S. Mancini, \emph{Insights
into the softening of chaotic statistical models by quantum considerations},
in Bayesian Inference and Maximum Entropy Methods in Science and Engineering,
AIP Conf. Proc. \textbf{1443}, 366 (2012).

\bibitem {PR4}S. A. Ali, C. Cafaro, A. Giffin, C. Lupo, and S. Mancini,
\emph{On a differential geometric viewpoint of Jaynes' MaxEnt method and its
quantum extension}, in Bayesian Inference and Maximum Entropy Methods in
Science and Engineering, AIP Conf. Proc. \textbf{1443}, 120 (2012).

\bibitem {PR2}A. Giffin, S. A. Ali, and C. Cafaro, \emph{Local softening of
chaotic statistical models with quantum consideration}, in Bayesian Inference
and Maximum Entropy Methods in Science and Engineering, AIP Conf. Proc.
\textbf{1553}, 238 (2013).

\bibitem {OSID}C. Cafaro, A. Giffin, C. Lupo and S. Mancini, \emph{Softening
the complexity of entropic motion on curved statistical manifolds}, Open Syst.
\& Inf. Dyn. \textbf{19}, 1250001 (2012).

\bibitem {SoftChaos}A. Giffin, S. A. Ali, and C. Cafaro,\emph{ Local softening
of information geometric indicators of chaos in statistical modeling in the
presence of quantum-like considerations}, Entropy \textbf{15}, 4622 (2013).

\bibitem {carlo-CSF}C. Cafaro, \emph{Works on an information geometrodynamical
approach to chaos}, Chaos, Solitons \& Fractals \textbf{41}, 886 (2009).

\bibitem {EJTP}C. Cafaro and S. A. Ali, \emph{Geometrodynamics of information
on curved statistical manifolds and its applications to chaos}, EJTP
\textbf{5}, 139 (2008).

\bibitem {zurek1}W. H. Zurek and J. P. Paz, \emph{Decoherence, chaos, and the
second law}, Phys. Rev. Lett. \textbf{72}, 2508 (1994); \emph{Quantum chaos: a
decoherent definition}, Physica \textbf{D83}, 300 (1995).

\bibitem {zurek2}W. H. Zurek, \emph{Preferred states, predictability,
classicality and environment-induced decoherence}, Prog. Theor. Phys.
\textbf{89}, 281 (1993).

\bibitem {carloMPLB}C. Cafaro, \emph{Information geometry, inference methods
and chaotic energy levels statistics}, Mod. Phys. Lett. \textbf{B22}, 1879 (2008).

\bibitem {cafaroPA}C. Cafaro and S. A. Ali, \emph{Can chaotic quantum energy
levels statistics be characterized using information geometry and inference
methods?}, Physica \textbf{A387}, 6876 (2008).

\bibitem {PR5}D.-H. Kim, S. A. Ali, C. Cafaro, and S. Mancini, \emph{An
information geometric analysis of entangled continuous variable quantum
systems}, Journal of Physics: Conference Series \textbf{306}, 012063 (2011).

\bibitem {kim2}D.-H. Kim, S. A. Ali, C. Cafaro and S. Mancini,
\emph{Information geometric modeling of scattering induced quantum
entanglement}, Phys. Lett. \textbf{A375}, 2868 (2011).

\bibitem {kim1}D.-H. Kim, S. A. Ali, C. Cafaro and S. Mancini,
\emph{Information geometry of quantum entangled Gaussian wave-packets},
Physica \textbf{A391}, 4517 (2012).

\bibitem {epr}A. Einstein, B. Podolsky, N. Rosen, \emph{Can quantum-mechanical
description of physical reality be considered complete?}, Phys. Rev.
\textbf{47, }777 (1935).

\bibitem {Schrodinger}E. Schr\"{o}dinger, \emph{Die gegenwartige situation in
der quantenmechanik}, Naturwissenschaften \textbf{23}: pp. 807-812; 823-828;
844-849 (1935).

\bibitem {Harshman1}N. L. Harshman and G. Hutton, $\emph{Entanglement}$
$\emph{generation}$ $\emph{in}$ $\emph{the}$ $\emph{scattering}$ $\emph{of}$
$\emph{one}$-$\emph{dimensional}$ $\emph{particles}$, Phys. Rev. \textbf{A77},
042310 (2008).

\bibitem {Harshman2}N. L. Harshman and P. Singh, \emph{Entanglement mechanisms
in one-dimensional potential scattering}, J. Phys. \textbf{A}: Math. Theor.
\textbf{41}, 155304 (2008).

\bibitem {mosca}P. Kaye, R. Laflamme, and M. Mosca, \emph{An Introduction to
Quantum Computing}, Oxford University Press (2007).

\bibitem {Wang}J. Wang, C. K. Law and M.-C. Chu, \emph{Loss of purity by
wave-packet scattering at low energies}, Phys. Rev. \textbf{A73}, 034302 (2006).

\bibitem {Law}C. K. Law, \emph{Entanglement production in colliding wave
packets}, Phys. Rev. \textbf{A70}, 062311 (2004).

\bibitem {Landau}L. D. Landau and E. M. Lifshitz, \emph{Quantum mechanics:
Non-relativistic theory}, Butterworth-Heinemann (1981).

\bibitem {Serafini}A. Serafini and G. Adesso, \emph{Standard forms and
entanglement engineering of multimode Gaussian states under local operations},
J. Phys. \textbf{A40}, 8041 (2007). V. M. Alekseev and M. V. Yakobson,
\emph{Symbolic dynamics and hyperbolic dynamic systems}, Phys. Reports
\textbf{75}, 287 (1981).

\bibitem {Mishima}K. Mishima, M. Hayashi and S. H. Lin, \emph{Entanglement in
scattering processes}, Phys. Lett. \textbf{A333}, 371 (2004).

\bibitem {Ropotenko}K. Ropotenko,\ \emph{Kolmogorov-Sinai entropy and black
holes}, Class. Quant. Grav. \textbf{25}; 195005 (2008).

\bibitem {Prosen}T. Prosen, M. Znidaric, \emph{Is the efficiency of classical
simulations of quantum dynamics related to integrability?} Phys. Rev.
\textbf{E75, }015202 (2007); T. Prosen, I. Pizorn, \emph{Operator space
entanglement entropy in transverse Ising chain}, Phys. Rev. \textbf{A76,
}032316\textbf{ }(2007).

\bibitem {nc}M. Gu, K. Wiesner, E. Rieper and V. Vedral, \emph{Quantum
mechanics can reduce the complexity of classical models}, Nature Commun.
\textbf{3}, 1 (2012).

\bibitem {beck}C. Beck and F. Schlogl, \emph{Thermodynamic Analysis of Chaotic
Systems: An Introduction}, Cambridge University Press (1995).

\bibitem {peng}L. Peng, H. Sun and G. Xu, \emph{Information geometric
characterization of the complexity of fractional Brownian motion}, J. Math.
Phys. \textbf{53}, 123305 (2012).

\bibitem {peng2}L. Peng, H. Sun, D.\ Sun and J. Yi, \emph{The geometric
structures and instability of entropic dynamical models}, Adv. Math.
\textbf{227}, 459 (2011).

\bibitem {r1}O. Semarak and P. Sukova, \emph{Free motion around black holes
with discs or rings: between integrability and chaos-I}, Monthly Notices of
the Royal Astronomical Society \textbf{404}, 545 (2010).

\bibitem {r2}C. Li, H. Sun and S. Zhang, \emph{Characterization of the
complexity of an ED model via information geometry}, Eur. Phys. J. Plus
\textbf{128}, 70 (2013).

\bibitem {r3}L. Cao, D. Li, E. Zhang, Z. Zhang and H. Sun, \emph{A statistical
cohomogeneity one metric on the upper plane with constant negative curvature},
Adv. Math. Phys., Vol. 2014 (2014), Article ID 832683, 6 pages.

\bibitem {r5}S. M. Abtahi, S. H. Sadati and H. Salarieh, \emph{Ricci-based
chaos analysis for roto-translatory motion of a Kelvin-type gyrostat
satellite}, Journal of Multi-Body Dynamics \textbf{228}, 34 (2014).

\bibitem {r6}J. Mikes and E. Stepanova,\emph{\ A five-dimensional Riemannian
manifold with an irreducible SO (3)-structure as a model of abstract
statistical manifold}, Annals of Global Analysis and Geometry \textbf{45}, 111 (2014).

\bibitem {r7}S. Weis, \emph{Continuity of the maximum-entropy inference},
Commun. Math. Phys. \textbf{330}, 1263 (2014).

\bibitem {r8}C. Li, L. Peng and H.\ Sun, \emph{Entropic dynamical models with
unstable Jacobi fields}, Rom. Journ. Phys. \textbf{60}, 1249 (2015).

\bibitem {r9}M. Itoh and H. Satoh, \emph{Geometry of Fisher information metric
and the barycenter map}, Entropy \textbf{17}, 1814 (2015).

\bibitem {r10}R. Franzosi, D. Felice, S. Mancini and M. Pettini, \emph{A
geometric entropy detecting the Erd\"{o}s-R\'{e}nyi phase transition}, Eur.
Phys. Lett. \textbf{111}, 20001 (2015).

\bibitem {r11}A. C. R. Martins, \emph{Opinion particles: Classical physics and
opinion dynamics}, Phys. Lett. \textbf{A379}, 89 (2015).

\bibitem {r12}S. A. Muhammad, E. Zhang and H.\ Sun, \emph{Jacobi fields on the
manifold of Freund}, Italian Journal of Pure and Applied Mathematics
\textbf{34}, 181 (2015).

\bibitem {r13}D. Felice and S. Mancini,\emph{\ Gaussian network's dynamics
reflected into geometric entropy}, Entropy \textbf{17}, 5660 (2015).

\bibitem {r14}C. Wen-Haw, \emph{A review of geometric mean of positive
definite matrices}, British Journal of Mathematics and Computer Science
\textbf{5}, 1 (2015).

\bibitem {r15}S. Weis, A. Knauf, N. Ay and M.-J. Zhao, \emph{Maximizing the
divergence from a hierarchical model of quantum states}, Open Syst. \& Inf.
Dyn. \textbf{22}, 1550006 (2015).

\bibitem {r16}S. Weis, \emph{Maximum-entropy inference and inverse continuity
of the numerical range}, Reports on Mathematical Physics \textbf{77}, 251 (2016).

\bibitem {r17}D. S. Shalymov and A. L. Fradkov, \emph{Dynamics of
non-stationary processes that follow the maximum of the R\'{e}nyi entropy
principle}, Proc. R. Soc. \textbf{A472}, 20150324 (2016).

\bibitem {r18}I. S. Gomez and M. Portesi, \emph{Ergodic statistical models:
entropic dynamics and chaos}, AIP Conf. Proc. \textbf{1853}, Art. ID:100001 (2017).

\bibitem {r18a}I. S. Gomez, \emph{Notions of the ergodic hierarchy for curved
statistical manifolds}, Physica \textbf{A484}, 117 (2017).

\bibitem {r19}G. Henry and D. Rodriguez, \emph{On the instability of two
entropic dynamical models}, Chaos,\ Solitons \& Fractals \textbf{91}, 604 (2016).
\end{thebibliography}
\end{document}